\documentclass[11pt, a4paper]{article}
\usepackage{jcappub}
\usepackage{natbib}
\usepackage{aas_macros}

\usepackage[T1]{fontenc}
\usepackage[utf8]{inputenc}
\usepackage[english]{babel}
\usepackage{amsmath}
\usepackage{amssymb}
\usepackage{bm}

\usepackage{graphicx}
\usepackage{subfig}
\usepackage{enumerate}
\usepackage{epsfig,epstopdf}
\usepackage{color,xcolor}
\usepackage{multirow}
\usepackage{mleftright}
\usepackage{soul}

\usepackage{url}
\usepackage{xspace}
\usepackage{comment}
\usepackage{bm}
\usepackage{braket}
\usepackage[noabbrev]{cleveref} 
\usepackage{hhline}

\usepackage{hyperref}
\newcommand{\kpar}{k_{\parallel}}
\newcommand{\kparmin}{k_{\parallel, {\rm min}}}
\newcommand{\kperp}{k_{\bot}}

\newcommand{\veps}{\varepsilon}
\newcommand{\gCDM}{$\gamma {\rm CDM}\,$}
\newcommand{\wCDM}{$w_0w_a{\rm CDM}\,$}
\newcommand{\bk}{\bm{k}}
\newcommand{\bq}{\bm{q}}
\newcommand{\bx}{\bm{x}}

\long\def\comment#1{}

\newcommand{\Mpc}{\ensuremath{\text{$h$/Mpc}}\xspace}
\newcommand*\mean[1]{\overline{#1}}

\newcommand*{\ie} {i.\,\!e.\;}
\newcommand*{\eg} {e.\,\!g.\;}

\newcommand*{\perm} {\;\text{perm}}
\newcommand{\Planck}{\textit{Planck}\,}

\graphicspath{{./plots/}}

\title{{Cosmological constraints from the power spectrum and bispectrum of 21cm intensity maps}}

\author[1]{Dionysios Karagiannis,}
\author[1,2,3]{Roy Maartens,}
\author[1]{Liantsoa F. Randrianjanahary}
\affiliation[1]{Department of Physics $\&$ Astronomy, University of the Western Cape, Cape Town 7535, South Africa}
\affiliation[2]{National Institute for Theoretical $\&$ Computational Sciences (NITheCS), Cape Town 7535, South Africa}
\affiliation[3]{Institute of Cosmology $\&$ Gravitation, University of Portsmouth, Portsmouth PO1 3FX, United Kingdom}

\emailAdd{dakaragian@gmail.com}

\abstract{
The 21cm emission of neutral hydrogen is a potential probe of the matter distribution in the Universe after reionisation. Cosmological surveys of this line intensity will be conducted in the coming years by the SKAO and HIRAX experiments, complementary to  upcoming galaxy surveys. We present the first forecasts of the cosmological constraints
from the combination of the 21cm power spectrum and bispectrum. Fisher forecasts are computed for the constraining power of these surveys on  cosmological parameters, the BAO distance functions and the growth function. We also estimate the constraining power on dynamical dark energy and modified gravity. Finally we investigate the constraints on the 21cm clustering bias, up to second order. We take into account the effects  on the 21cm correlators of the telescope beam, instrumental noise and foreground avoidance, as well as the Alcock-Paczynski effect and the effects of theoretical errors in the modelling of the correlators. 
We find that, together with \Planck priors, and marginalising over clustering bias and nuisance parameters, HIRAX achieves sub-percent precision on the $\Lambda$CDM parameters, with SKAO delivering slightly lower precision. The modified gravity parameter $\gamma$ is constrained at 1\% (HIRAX) and 5\% (SKAO). For the dark energy parameters $w_0,w_a$, HIRAX delivers percent-level precision while SKAO constraints are weaker.  HIRAX achieves sub-percent precision on  the BAO distance functions $D_A,H$,  while SKAO reaches $1-2\%$ 
for $0.6\lesssim z\lesssim 1$. The growth rate $f$ is constrained at a few-percent level for the whole redshift range of HIRAX and for $0.6\lesssim z\lesssim 1$ by SKAO. 
The different performances arise mainly since HIRAX is a packed inteferometer that is optimised for BAO measurements, while SKAO is not optimised for interferometer cosmology and operates better in single-dish mode, where the telescope beam limits access to the smaller scales that are covered by an interferometer.
}

\begin{document}

\maketitle
\clearpage
\section{Introduction}

The tightest constraints on cosmological parameters have been provided by cosmic microwave background (CMB) anisotropies as measured by \Planck \cite{Planck2018_cosmo} and by large-scale structure (LSS) surveys such as SDSS and DES. Next-generation LSS surveys with DESI \cite{DESI2016}, Euclid \cite{Euclid:2019clj}, LSST \cite{LSSTDarkEnergyScience:2018jkl}, SKAO \cite{Bacon:2018dui} and HIRAX \cite{Crichton:2021hlc}, combined with CMB data, will significantly improve over the current measurements. 

The post-reionisation 21cm emission line of neutral hydrogen (HI) is a tracer of the underlying matter distribution. Detecting individual 21cm-emitting galaxies is a difficult task, especially at higher redshifts, given the weakness of the line. If instead we measure the integrated emission in each pixel, we can perform large-volume surveys of the HI fluctuations -- known as 21cm (equivalently, HI) intensity mapping surveys \cite{Santos:2015gra}. Such spectroscopic surveys are planned with SKAO-MID (hereafter referred to as SKAO) and HIRAX, which we  consider here. They will provide an exciting and important complementary probe to the traditional optical/ infra-red LSS surveys, with completely different systematics. In addition to their wide sky areas, these surveys will together cover the redshift range $0<z\lesssim 3$, beyond the range accessed by cosmological galaxy surveys.

Cosmological constraints are typically performed using the power spectrum. The addition of the bispectrum is known to improve constraining power (to varying degrees) and to break parameter degeneracies, as has been shown in recent work on BOSS galaxy surveys \cite{Ivanov:2021kcd,Philcox:2021kcw,DAmico:2022gki,Cabass:2022ymb}. (For recent work on the galaxy bispectrum, see e.g. \cite{Veropalumbo:2022zfs,Rizzo:2022lmh,Coulton:2022qbc,Jung:2022rtn,Philcox:2022frc}.) 
Regarding 21cm intensity mapping, the combination of the power spectrum and bispectrum has been used to investigate future constraints on primordial non-Gaussianity 
\cite{Karagiannis:2019jjx,Karagiannis:2020dpq}. Here we use the same combination to forecast  constraints from 21cm intensity mapping surveys on standard cosmological parameters, dark energy parameters and a modified gravity parameter. We also derive constraints on redshift-dependent BAO (baryon acoustic oscillation) distances and on the growth rate function.

In contrast to galaxy surveys, HI intensity mapping is contaminated by huge foregrounds, similar to the CMB. Within the simplified framework of Fisher forecasting, we follow the usual approach of foreground avoidance (e.g. \cite{Bull2015,Karagiannis:2019jjx}), since foreground cleaning requires substantial numerical simulations (e.g. \cite{Spinelli:2021emp}). Another difference lies in the noise: for HI intensity mapping on linear/ quasi-linear scales, the shot noise can be neglected since it is dominated by instrumental noise. Apart from these two differences we apply a fairly standard analysis, based on the tree-level power spectrum and bispectrum, but with modifications to incorporate nonlinear redshift-space distortions, which affect smaller scales than the tree-level limit. We also apply `theoretical errors' to the covariances in order to take account of inaccuracies in the modeling. In our Fisher analysis, we marginalise over all nuisance parameters.

Our overall finding is that the upcoming HI intensity mapping surveys can deliver exquisite precision on the cosmological parameters and functions, as well as on the clustering bias parameters, with HIRAX out-performing SKAO, based on the different designs of the two surveys.

The remainder of the paper is organised as follows.
\Cref{sec:model_PK_BK} presents the theoretical analysis of the redshift-space power spectrum and bispectrum, including the clustering biases and the 
Alcock-Paczynski (AP) effect. Details of the HI intensity mapping surveys are presented in \Cref{sec:surveys}, including the telescope specifications, the associated instrumental noise and the effect of the telescope beam. We make clear the differences between interferometer surveys (HIRAX and its precursor) and single-dish surveys (SKAO and its precursor). In \Cref{method} we discuss in some detail the Fisher analysis and the various parameters that are included with the relevant priors. The theoretical error contribution to the covariances is also summarised. Our results are presented in \Cref{sec:results}, in the form of tables, contour plots and plots of redshift-dependent errors. We also interpret these results and comment on the differences between the interferometer and single-dish surveys. Finally, \Cref{sec:conc} summarises our main results.

\section{Theoretical Model}\label{sec:model_PK_BK}

The HI power spectrum and bispectrum in redshift space will be modelled perturbatively by using the Standard Perturbation Theory (SPT), which assumes that the dynamics of long-wavelength density and velocity perturbations are driven by the hydrodynamics of an Eulerian pressureless perfect fluid (see \cite{Bernardeau2002} for a review). Moreover, a complete bias and redshift space distortions (RSD) expansion, up to the lowest non-vanishing order in the perturbations, will be considered. A phenomenological non-perturbative description will be used for the so-called fingers-of-God effect (FoG) \cite{Jackson1972}, caused by the virialized motions of galaxies, which cannot be described within the framework of SPT. Gaussian initial conditions will be considered throughout this work, while the analysis will be restricted within the linear/semi-linear regime, where the tree-level power spectrum and bispectrum provide an adequate description.

\subsection{Matter power spectrum and bispectrum}\label{sec:matter_PK_BK}

The power spectrum of the Bardeen gauge-invariant primordial gravitational potential is defined in Fourier space by
\begin{equation}
 \langle \Phi(\bk)\Phi(\bk')\rangle=(2\pi)^3 \delta_{\rm D}(\bk+\bk')P_\Phi(k),  
\end{equation}
where $P_\Phi(k)$ is related to the power spectrum of the primordial curvature perturbations, generated during inflation. For the standard single-field slow-roll inflationary scenario, their distribution should be nearly perfect Gaussian. The primordial perturbations $\Phi$ are in turn related to the linear dark matter density contrast through the Poisson equation, $\delta_{\rm m}^{\rm L}(\bk,z)=M(k,z)\Phi(\bk)$, where
 \begin{eqnarray}
 \label{eq:poisM}
 M(k,z)&=&\frac{2c^2D(z)}{3\Omega_{\rm m} H_0^2\, }\,T(k)\, k^2.
 \end{eqnarray}
 Here $D(z)$ is the growth factor of the linearly evolved density contrast, normalised to unity today (\ie $D(0)=1$), and $T(k)$ is the matter transfer function normalized to unity at large scales, $k \rightarrow 0$. The linear matter power spectrum ($P_{\rm m}^L(k,z)=M^2(k,z)P_\Phi(k)$) will be computed with the numerical Boltzmann code CAMB \cite{CAMB}.
 
 For Gaussian initial conditions, higher-order correlators are non-zero due to the non-linearities induced by gravity. The most important is the bispectrum, \ie the Fourier transform of the three-point function:\footnote{We use the ordering $k_3\le k_2\le k_1$.}
 \begin{equation}
  \langle\delta_{\rm m}(\bk_1) \delta_{\rm m}(\bk_2) \delta_{\rm m}(\bk_3)\rangle=(2\pi)^3\delta_{\rm D}(\bk_1+\bk_2+\bk_3)B_{\rm m}(k_1,k_2,k_3).
 \end{equation}
 Here the Dirac delta function ensures the conservation of momentum. 
 
 The fiducial cosmology is given by the average values of the flat $\Lambda$CDM model, measured by the \Planck mission. In particular we use the {\tt base\_plikHM\_TTTEEE\_lowl\_lowE\_lensing} column of the \Planck 2018 results \cite{Planck2018_cosmo}.

 \subsection{Neutral hydrogen bias} \label{sec:halo_bias}


Cosmological forecasting from future HI IM surveys (see \Cref{sec:surveys} for details) requires a robust description of the relation between the statistics of observed tracers and the underlying distribution of dark matter, \ie the clustering {bias}. 

The density contrast of halos $\delta_{\rm h}$ can be expressed perturbatively as a series of operators, constructed out of all possible local gravitational observables \cite{Assassi2014,Senatore2014,Mirbabayi2014} (\ie operators formed by the tidal tensor $\partial_i\partial_j\Phi$ and its derivatives, where $\Phi$ can be the Newtonian gravitational potential or the velocity potential $\Phi_\upsilon$), satisfying rotational symmetry and the equivalence principle (see e.g.~\cite{Desjacques2016} for a review). For Gaussian initial conditions and the spacial scales considered here, the complete set of terms up to second order is needed. The Eulerian halo density overdensity can be written as  
  \begin{equation}\label{eq:deltaG}
   \delta_{\rm h}^{\rm E}(\bx,\tau)= b_1^{\rm E}(\tau)\delta_{\rm m}(\bx,\tau) +\veps^{\rm E}(\bx,\tau)+\frac{b_2^{\rm E}(\tau)}{2} \delta_{\rm m}^2(\bx,\tau) + \frac{b_{s^2}^{\rm E}(\tau)}{2}s^2(\bx,\tau)+\veps_{\delta}^{\rm E}(\bx,\tau)\delta_{\rm m}(\bx,\tau) \,,
  \end{equation}
  where $\tau$ is conformal time, $\bx$ are  spatial comoving coordinates in the Eulerian frame, $s^2=s_{ij}s^{ij}$
  is the simplest scalar that can be formed from the tidal field, $\veps^{\rm E}$ is the leading stochastic field \cite{Dekel1998,Taruya1998,Matsubara1999} and $\veps_{\delta}^{\rm E}$ is the stochastic field associated with the linear bias. These fields take into account the stochastic relation between the galaxy density and any large-scale field. The higher-order derivative term, which encapsulates short-scale dynamics and is present at second-order, is excluded from the expansion since the spacial scales considered here are much larger than the Langrangian radius of halos hosting the galaxies of interest.

 Applying the general bias expansion, described before, to the neutral hydrogen, an additional ingredient is needed, namely the description on how HI is distributed within the dark matter halos. This can be achieved within the framework of the halo model \cite{Seljak2000,Peacock2000,Scoccimarro2000}. In this approach, HI is assumed to occupy regions within the halos, with a negligible contribution outside of them. The HI density is then defined as \cite{Villaescusa2014,Castorina2016}: 
  \begin{equation}
      \rho_{\rm HI}(z)=\int {\rm d} M\, n_{\rm h}(M,z) M_{\rm HI}(M,z) ,
  \end{equation}
  where $M_{\rm HI}$ is the average HI mass within the halo of total mass $M$ at redshift $z$. The halo mass function,  $n_{\rm h}(M,z)$, is considered to be the best-fit results of \cite{Tinker2008}, which originate from fitting to N-body simulations. For $M_{\rm HI}$ we use a halo occupation distribution (HOD) approach \cite{Cooray2002} and follow the model of \cite{Castorina2016}:  
  \begin{equation}\label{eq:HOD}
    M_{\rm HI}(M,z)=C(z)(1-Y_p)\frac{\Omega_{\rm b}}{\Omega_{\rm m}}\,{\rm e}^{-M_{\rm min}(z)/M}\,M^{q(z)},
  \end{equation}
  where $C$ is a normalization constant,  $Y_p = 0.24$ is the helium fraction, $M_{\rm min}$ is the halo mass below which the amount of HI in halos is exponentially suppressed,  and $q$ controls the efficiency of generating or destroying neutral hydrogen inside halos. The power-law is in agreement with the numerical results from hydrodynamic simulations of~\cite{Villaescusa-Navarro:2015cca,Villaescusa-Navarro:2015zaa}, while the presence of the exponential cut-off ensures the suppression of HI in low-mass halos \cite{Pontzen:2008mx,Marin2010,Villaescusa2014}. The fiducial values of the HOD free parameters are ${q}=1$ and $M_{\rm min}=5\times 10^9 M_{\odot}/h$.

 The HI bias coefficients are given by  
  \begin{equation}\label{eq:bHI}
      b_{\rm HI}^i(z)=\frac{1}{\rho_{\rm HI}(z)}\int_0^\infty  {\rm d} M\, n_{\rm h}(M,z) b_i^{\rm h}(M,z)M_{\rm HI}(M,z),
  \end{equation}
   where the index $i$ corresponds to the subscripts of the bias terms in the expansion of \Cref{eq:deltaG}. For the linear halo bias we use the the fitting function of~\cite{Tinker2010}, while for the quadratic bias the analytic expression is derived after using the mass function of \cite{Tinker2008} and the peak-background split argument \cite{Lazeyras2016,Karagiannis:2019jjx}. In~\cite{Lazeyras2016} it is shown that both expressions are in good agreement with numerical results for the HI mass ranges considered here. The second-order tidal field bias coefficient is related to the linear bias by $b_{s^2}^{\rm E}=-4(b_1^{\rm E}-1)/7$ \cite{Baldauf2012}.

\subsection{Power spectrum and bispectrum in redshift space }\label{sec:RSDmodel}

The distance of a luminous object is determined by its motion with the Hubble flow, which is affected by its peculiar velocity. This effect is known as a redshift space distortion \cite{Sargent:1977,Kaiser1987,Hamilton1998} and can be taken into account by mapping the real space correlator to redshift space. Here we will consider the flat-sky approximation, \ie the line-of-sight vector $\hat{\bm z}$ is a constant unit vector. In the non-perturbative regime, the velocity dispersion of objects during the virialisation process make structures appear more elongated along the line of sight, compared to real space, \ie the FoG effect. It is treated here phenomenologically, by introducing an exponential damping factor, which models the suppression of  clustering power in redshift space.

The tree-level expressions for the HI power spectrum and  bispectrum in redshift space are given by:
     \begin{align}
     P_{\rm HI}(\bk,z)&=\bar{T}(z)^2\Big[D_\text{FoG}^P(\bk,z)Z_1(\bk,z)^2P_{\rm m}(k,z)+P_{\veps}(z)\Big]+P_{\rm N}(\bk,z) \label{eq:Pgs},\\ 
   B_{\rm HI}(\bk_1,\bk_2,\bk_3,z)&= \bar{T}(z)^3\bigg\{D_\text{FoG}^B(\bk_1,\bk_2,\bk_3,z) \nonumber \\
   &~\times\bigg[2Z_1(\bk_1,z)Z_1(\bk_2,z)Z_2(\bk_1,\bk_2,z)P_{\rm m}^{\rm L}(k_1,z)P_{\rm m}^{\rm L}(k_2,z)+2~ \text{perm}\bigg] \nonumber \\
   &~+2P_{\veps\veps_{\delta}}(z)\Big[Z_1(\bk_1,z)P_{\rm m}^{\rm L}(k_1,z)+2~ \text{perm}\Big]+B_{\veps}(z)\bigg\}. \label{eq:Bgs} 
  \end{align}
In HI IM, $P_{\rm N}$ is the instrumental noise (see \Cref{sec:surveys}), {where it is assumed to be Gaussian (see \cite{Bull2015} for a discussion) and therefore it is only present in the two-point correlator}. The background temperature is $\bar{T}=188\,\Omega_{\rm HI} (z)h(1+z)^{2}H_0/H(z)~\mu$K, where the cosmic evolution of the HI density is modelled as $\Omega_{\rm HI}(z)=4\times 10^{-4}(1+z)^{0.6}$ \cite{Battye2013}. 
The general  redshift kernels up to second order and for Gaussian initial conditions are:
  \begin{align}
   &Z_1(\bk_i)=b_1+f\mu_i^2, \label{eq:Z1}\\
   &Z_2(\bk_i,\bk_j)=b_1F_2(\bk_i,\bk_j)+f\mu_{ij}^2G_2(\bk_i,\bk_j)+\frac{b_2}{2} +\frac{b_{s^2}}{2}S_2(\bk_i,\bk_j) 
   \nonumber \\ &
   +\frac{1}{2}f\mu_{ij}k_{ij}\left[\frac{\mu_i}{k_i}Z_1(\bk_j)+\frac{\mu_j}{k_j}Z_1(\bk_i)\right], \label{eq:Z2}
  \end{align}
where $f$ is the linear growth rate, $\mu_i=\hat\bk_i\cdot\hat{\bm{z}}$, $\mu_{ij}=(\mu_ik_i+\mu_jk_j)/k_{ij}$ and $k_{ij}^2=(\bk_i+\bk_j)^2$. Note that we  suppressed the redshift dependence for brevity. The kernels $F_2(\bk_i,\bk_j)$ and $G_2(\bk_i,\bk_j)$ are the second-order symmetric SPT kernels \cite{Bernardeau2002}, while $S_2(\bk_1,\bk_2) = (\hat\bk_1\cdot\hat\bk_2)^2-1/3$ is the  tidal kernel \cite{McDonald2009,Baldauf2012}. The FoG damping factors are \cite{Peacock1994,Ballinger1996}
  \begin{align}
 D_\text{FoG}^P(\bk,z)&=\exp\big[-k^2\mu^2\,\sigma_P(z)^2\big], \label{eq:fog_PS}\\
 D_\text{FoG}^B(\bk_1,\bk_2,\bk_3,z)&=\exp\big[-\big(k_1^2\mu_1^2+k_2^2\mu_2^2+k_3^2\mu_3^2\big)\sigma_B(z)^2\big] \label{eq:fog_BS},
  \end{align}
  where the damping parameters $\sigma_P$ and $\sigma_B$ have fiducial value equal to the linear  velocity dispersion $\sigma_\upsilon$. The stochastic terms (\ie $P_{\veps}$, $P_{\veps\veps_{\delta}}$ and $B_{\veps}$) approach their asymptotic constant values as $k\rightarrow 0$. For the scales considered here, the Poisson distribution characterises fully the correlation of these components, hence these predictions will be used as fiducial values for the shot-noise contributions. In the HI halo model formalism (\Cref{sec:halo_bias}), the shot noise term is given by \cite{Castorina2016}
  \begin{equation} \label{eq:IM_shot}
      P_{\rm SN}(z)=\frac{1}{\mean{n}_{\rm eff}(z)}=\frac{1}{\rho_{\rm HI}(z)}\int {\rm d}\ln M\, n_{\rm h}(M,z)M_{\rm HI}^2.
  \end{equation}
 The effective number density can in turn be used for the fiducial values of all the stochastic contributions \cite{Schmidt2015,Desjacques2016}:
  \begin{equation}\label{eq:poisson_fid}
   P_{\veps}\equiv P_{\rm SN}, \quad P_{\veps\veps_{\delta}}=\frac{b_1}{2\mean{n}_{\rm eff}}, \quad B_{\veps}=\frac{1}{\mean{n}_{\rm eff}^2}.
  \end{equation}

The redshift-space bispectrum is characterized by five variables: three to define the triangle shape (e.g. the
sides $k_1$, $k_2$, $k_3$) and  two to characterize the orientation of the  triangle relative to the line of sight. The angles characterising this orientation are, following \cite{Scoccimarro1999}, the polar angle $\theta$ of $\bk_1$, with $\cos \theta=\mu\equiv\hat{\bk}_1\cdot\hat{\bm z}$ and the azimuthal angle $\phi$ around $\bk_1$. Then the cosines of the angles that the wave-vectors make with the line of sight are: $\mu_1=\mu$, $\mu_2=\mu\cos x_{12}+\sqrt{1-\mu^2}\sin x_{12}\sin\phi$ and $\mu_3=-(k_1/k_3)\mu_1-(k_2/k_3)\mu_2$, where  $\cos x_{12}=\hat{\bk}_1\cdot\hat{\bk}_2$. Then  $B_{\rm HI}(\bk_1,\bk_2,\bk_3,z)=B_{\rm HI}(k_1,k_2,k_3,\mu_1,\phi,z)$. Here we focus on the monopole of the tree-level bispectrum, obtained after taking the average over all directions. In \cite{Gagrani:2016rfy}, it is shown that the lowest-order bispectrum multipoles do not suffer from a significant loss of information on most of the cosmological parameters and bias coefficients (see also \cite{Yankelevich2018} for a discussion).

  In this work we stay mostly within the perturbative regime, slightly venturing into mildly nonlinear scales for low redshift. For the power spectrum, the tree-level description is sufficient. Nonetheless, in order to improve the precision of the matter modelling, instead of using the linear power spectrum $P^{\rm L}_{\rm m}$ to describe $P_{\rm m}$ in \Cref{eq:Pgs}, we  use the non-linear matter power spectrum, $P^{\rm NL}_{\rm m}$, from the updated version of the HMCode augmented halo model \cite{Mead2020}. The HMCode approach provides more accurate predictions, over a large range of scales, relative to the usual Halofit model \cite{Smith2003,Takahashi2012}, as shown in \cite{Smith:2018zcj,Mead2020}. In addition, a general feature of the Halofit approach is the poor performance in predicting the derivatives of the power spectrum with respect to some cosmological parameters \cite{Reimberg2018}. This indicates that the usage of Halofit, in a Fisher matrix error forecast, hides inaccuracies and should be used with extreme caution. In order to avoid the possibility of unreliable Fisher information matrix calculation, we use the HMCode, evaluated with the latest version\footnote{\url{https://camb.readthedocs.io/en/latest/}}  of CAMB, to describe the non-linear matter power spectrum. 
  
  For the bispectrum, the tree-level modelling provides an adequate description for the high-order clustering of HI on the scales considered here \cite{Gil-Marin:2014sta,Lazanu2015b,Hashimoto:2017klo,Chan2017,Oddo:2019run,Agarwal:2020lov,MoradinezhadDizgah2020,Ivanov:2021kcd}. Nonetheless, for both correlators, we will take into account the parameter shift due the exclusion of higher-order contributions in the matter and bias expansions, at the Fisher matrix level, through the `theoretical errors' approach (see \Cref{sec:theoretical_er}).

  \subsection{Alcock-Paczynski effect}\label{sec:APeffect}
  
  Another source of anisotropies in the observed galaxy clustering, in addition to RSD,  is the AP effect \cite{Alcock1979}, which occurs when the fiducial cosmology, used to convert the observed angular coordinates and redshifts to physical distances, differs from the true one. This leads to an artificial anisotropic distortion of the inferred galaxy distribution, which modulates the amplitude and shape of the power spectrum and bispectrum. The radial and transverse distortions are proportional to the Hubble parameter $H(z)$ and angular diameter distance $D_A(z)$ respectively. This provides additional information on probing the underlying cosmology, and this effect is taken into account in our forecasts. 
  
  The AP distortions rescale the radial and transverse components of $\bk$ (fiducial cosmology) as, $q_\parallel=\kpar [H_{\rm true}(z)/H_{\rm fid}(z)]$ and $q_\bot =\kperp [D_{A,\rm fid}(z)/D_{A,\rm true}(z)]$, where $\bq$ is the wave-vector in the true cosmology. The relation between the fiducial  ($k$, $\mu$) and the true ($q$, $\nu$) is given by:
\begin{align}
       q(k,\mu)&=k\alpha(\mu)\;, \quad
       \nu(k,\mu)=\frac{\mu}{\alpha(\mu)}\frac{H_{\rm true}}{H_{\rm fid}}\;, \label{eq:nu2mu}\\
 \alpha(\mu)&=\left[\left(\frac{H_{\rm true}}{H_{\rm fid}}\right)^2\mu^2+\left(\frac{D_{A,\rm fid}}{D_{A, \rm true}}\right)^2\left(1-\mu^2\right)\right]^{1/2},  
 \end{align}
   where here and below we suppress the redshift dependence for clarity. The observed power spectrum with AP effect \cite{Seo2003} and bispectrum with AP effect \cite{Song2015} are
        \begin{align} \label{eq:PS_AP}
       P^{\rm obs}_{\rm HI}\left(k,\mu,z\right) &=\left(\frac{H_{\rm true}}{H_{\rm fid}}\right)\left(\frac{D_{A,\rm fid}}{D_{A, \rm true}}\right)^2P_{\rm HI}\big(q,\nu,z \big) ,\\
    \label{eq:BS_AP}
       B^{\rm obs}_{\rm HI}\left(k_1,k_2,k_3,\mu_1,\phi,z\right)&=\left(\frac{H_{\rm true}}{H_{\rm fid}}\right)^2\left(\frac{D_{A,\rm fid}}{D_{A, \rm true}}\right)^4B_{\rm HI}\left(q_1,q_2,q_3,\nu_1,\phi,z\right). 
   \end{align}
   Note that for the Fisher matrix calculations, only $H_{\rm true}$ and $D_{A, \rm true}$ are varied (\ie free parameters), where their fiducial values are taken to be those that correspond to the fiducial cosmology, while $H_{\rm fid}$ and $D_{A, \rm fid}$  remain fixed.

  \section{HI intensity mapping surveys}\label{sec:surveys}

Radio telescopes can probe the Universe in two distinct ways: in interferometer (IF) mode, by correlating the signal from all dishes/dipole stations and outputting directly the Fourier transformation of the sky; or single-dish (SD) mode, by providing separate maps of the sky, which are added together to reduce noise, and the final summed map is then Fourier transformed.

The surveys considered here include current and near-future surveys: 
MeerKAT\footnote{\url{www.sarao.ac.za/science/meerkat/}}, a 64-dish already-operational precursor for SKAO\footnote{\url{www.skatelescope.org}} (which will have 64+133 dishes), functioning in SD mode, and HIRAX\footnote{\url{hirax.ukzn.ac.za}} \cite{Crichton:2021hlc}, which will have initially 256 dishes and then 1024, operating in IF mode. 

In the case of a line intensity mapping survey the noise component on relevant scales is dominated by the thermal noise from the instrument, while the shot-noise contribution remains minimal \cite{2011ApJ...740L..20G}. In IF mode, a Gaussian model of instrumental noise is given by \cite{Zaldarriaga2003b,Tegmark2008}
 \begin{equation}\label{eq:Pnoise_IF}
  P^{\rm IF}_{\rm N}(\kperp,{z})=T_{\rm sys}(z)^2\chi(z)^2\lambda(z)\frac{(1+z)}{H(z)}\left[\frac{\lambda(z)^2}{A_{\rm e}}\right]^2\frac{1}{2\,n_{\rm b}({u},z)\,t_{\rm{survey} }}\frac{S_{\rm area}}{ \theta_{\rm b}(z)^2}\,,
 \end{equation}
where  $t_{\rm{survey} }$ is the integration time, $S_{\rm area}$ is the survey sky area, and $\lambda(z)=\lambda_{21}(1+z)$ is the observed wavelength of the 21cm line. The field of view of a dish with diameter $D_{\rm dish}$ is  $\theta_{\rm b}^2$, where $\theta_{\rm b}(z)=1.22\,\lambda(z)/D_{\rm dish}$ is the FWHM of the beam of an individual dish. The effective area $A_{\rm e}=\eta\pi (D_{\rm dish}/2)^2$ depends on the efficiency $\eta$; for HIRAX we take $\eta=0.7$. The system temperature $T_{\rm sys}$ is the receiver temperature $T_{\rm rx}=50\,{\rm K}$ for HIRAX plus the sky temperature $T_{\rm sky}$, which is taken from Appendix D of \cite{PUMA2018}. The survey specifications are presented in \Cref{table:survey_specs_ITF}.

The  baseline density $n_{\rm b}$ is defined in the image plane, where we assume azimuthal symmetry. At an observed wavelength $\lambda$, the physical baseline corresponding to $u$ is $L=u\lambda$. $n_{\rm b}$  vanishes beyond the maximum baseline. For  HIRAX, we take the baseline distributions from simulations of the array\footnote{We thank Warren Naidoo for providing the simulated data used in \cite{Crichton:2021hlc}.}, 
 presented in \Cref{app:baseline}.

\begin{table}[t]
 \centering
 \begin{tabular}{l|c|c}
  IF Survey & HIRAX-256 & HIRAX-1024\\ \hline\hline
   redshift   &  $0.775-2.55$ & $0.775-2.55$  \\
 $N_{\rm dish}$ & $256$ & $1,024$\\
 $D_{\rm dish}$ [m] & $6$ & $6$ \\
  $D_{\rm max}$ [km] & $0.25$ & $0.25$ \\
 $S_{\rm area}$ [$\rm{deg}^2$] & $15,000$ & $15,000$ \\
 $t_{\rm survey}$ [hrs] & $17,500$ & $17,500$
 \end{tabular}
 \caption{Details for the interferometer-mode surveys with HIRAX  (from \cite{Crichton:2021hlc}).}
 \label{table:survey_specs_ITF}
\end{table}

\begin{table}[t]
 \centering
 \begin{tabular}{l|cc|cc}
  SD Survey & \multicolumn{2}{c|}{MeerKAT} & \multicolumn{2}{c}{SKAO$^{a}$} \\ \hline\hline
  & L Band & UHF Band & Band 1 & Band 2  \\ \hline 
   redshift   &  0.1$^{b}-$0.58 & 0.4$^{c}-$1.45 & 0.35$^{d}-$3.05 & 0.1$^{b}-$0.49  \\
 $N_{\rm dish}$ & $64$ & $64$ & $197$ & $197$ \\
 $D_{\rm dish}$ [m] & $13.5$ & $13.5$ & $15$ & $15$ \\
 $S_{\rm area}$ [$\rm{deg}^2$] & $4,000$ & $4,000$ & $20,000$ & $20,000$ \\
 $t_{\rm survey}$ [hrs] & $4,000$ & $4,000$ & $10,000$ & $10,000$
 \end{tabular}
 \caption{As in  \Cref{table:survey_specs_ITF}, for the single-dish mode surveys with MeerKAT (from \cite{Santos:2017qgq}) and SKAO (from \cite{Bacon:2018dui}). Notes: (a) The 64 MeerKAT dishes included in SKAO will keep their original specifications. For simplicity we neglect this difference and assume all dishes have SKAO specifications (see \cite{Fonseca:2019qek} for an accurate treatment). (b) Band covers redshift range $z=0-0.1$ which we neglect due to nonlinear effects. (c) $0.4\leq z\leq 0.58$ is excluded to avoid double-counting Fisher information in overlapping redshift bins, when using the available information from both bands. (d) As in (c), we exclude $0.35\leq z\leq 0.49$.}
 \label{table:survey_specs_SD}
\end{table}

For surveys in SD mode, the power spectrum of the instrumental noise is \cite{Santos:2015gra}:
\begin{equation}\label{eq:Pnoise_SD}
 P^{\rm SD}_{\rm N}(\kperp,z)=T_{\rm sys}(z)^2\chi(z)^2\lambda(z)\frac{(1+z)}{H(z)}\, \frac{S_{\rm area}}{\eta\, N_{\rm pol}\,N_{\rm dish}\,t_{\rm survey}\,{\beta_\perp}(\kperp,z)^2}  
 \,,
\end{equation}
where we assume that the dishes have a single feed and that the efficiency for both surveys is $\eta=1$, while the polarisation per feed is $N_{\rm pol}=2$. The system temperatures for MeerKAT and SKAO 
follow \cite{Bacon:2018dui}. The transverse effective beam is given in Fourier space by \cite{Bull2015}
\begin{equation}\label{eq:beam}
    \beta_\perp(k,\mu,z)=\exp\left[-\frac{k_\perp^2 \chi(z)^2 \theta_{\rm b}(z)^2}{16\ln 2} \right].
\end{equation}
 Due to the very high frequency resolution of IM experiments, the effective beam in the radial direction may be neglected \cite{Bull2015}. 

Cosmological survey specifications for MeerKAT are taken from \cite{Santos:2017qgq} and  for SKAO from \cite{Bacon:2018dui}, and are presented  in \Cref{table:survey_specs_SD}. Note that, in the analysis that follows, we assume that the full surveys will be done in each band for the two telescopes. The Fisher information matrix from the two bands of MeerKAT, as well as SKAO, will be added together, effectively treating MeerKAT and SKAO as two single-band surveys. To avoid double counting the information that lies within the overlapping redshift bins, we exclude redshifts within the range of $0.4-0.58$ in the case of the MeerKAT UHF-Band and $0.35-0.49$ for SKAO Band~1.

Note that the computation of thermal noise depends not only on the technical survey specifications, but also on the cosmological parameters via $H(z)$ and the comoving distance $\chi(z)$. 

The foreground emission from the Galaxy and astrophysical sources is orders of magnitude larger than the desired cosmological 21cm signal \cite{Shaw:2013wza,Shaw:2014khi,Pober:2014lva,Byrne:2018dkh,Spinelli:2021emp}. This effect contaminates the long-wavelength radial Fourier modes, so that modes with $k_\parallel<\kparmin$ are inaccessible \cite{Jacobson:2003wv,Furlanetto:2006jb,Chang2007,Liu2011,Liu2012,Shaw:2013wza,Shaw:2014khi}. The separation of the signal from the foreground emission is a great challenge. However, the  radio foregrounds are mainly very spectrally smooth free-free and synchrotron emission from our Galaxy and other unresolved sources. This characteristic makes possible the separation from the cosmological signal, which varies along the line-of-sight due to the underlying density field, without significant losses up to some small value of $\kparmin$ \cite{Liu2011,Liu2012,Shaw:2013wza,Shaw:2014khi}. 
Reconstruction techniques have been developed, to recover the long radial modes that are lost to foregrounds, by using the measured short modes. In the context of HI intensity mapping, this has been applied in \cite{Zhu:2016esh,Karacayli:2019iyd,Modi:2019hnu}, while in \cite{Jasche2010,Kitaura2013,Wang:2014hia,Jasche:2014vpa,Shaw:2014khi,Wang:2016qbz,Seljak:2017rmr,Modi:2018cfi} it shown that by using the forward model reconstruction framework, modes up to $\kpar\simeq 0.01\;\Mpc$ can be almost perfectly recovered.
   
We follow a foreground-avoidance approach and impose a hard cut-off on $\kpar$,  keeping in the Fisher analysis  only the  modes that satisfy: 
  \begin{equation}\label{eq:kparmin}
 \kpar \geq\kparmin \qquad\mbox{where}\qquad\kparmin=0.01\;\Mpc\,.
 \end{equation}
 In order to assess the effect of this foreground cut on constraining the parameters of interest, we also consider the idealised case
  $\kparmin=0$ and the less optimistic case $\kparmin=0.05\;\Mpc$ (see \Cref{sec:conc} for a discussion). 
 
 In the case of an interferometer, an additional instrumental effect arises  via the leakage of foregrounds to transverse modes, due to the chromatic response of the interferometer itself \cite{Liu2011,Liu2012,Parsons2012,Pober2014,Seo2015,Pober2015}. This is  not a fundamental astrophysical limitation, but a technical issue:  with excellent baseline-to-baseline calibration it can, in principle, be removed \cite{Seo2015}. Here we take the effect into account by excluding all modes lying in the `foreground wedge', \ie requiring that:
 \begin{equation}\label{eq:kwedge}
k_{\parallel}\geq A_{\rm wedge}(z)\,\kperp \,. 
 \end{equation}
The wedge factor $A_{\rm wedge}$ is determined by the source furthest from the zenith that can corrupt the data  \cite{Pober2014}:
 \begin{equation}\label{eq:wedge_prim}
  A_{\rm wedge}(z)=\frac{\chi(z)H(z)}{c(1+z)}\sin\big[ 0.61 N_{\rm w} \,\theta_{\rm b}(z)\big],
 \end{equation}
where sources up to $N_{\rm w}$ primary beam sizes away from the zenith can have an effect. We take
$N_{\rm w}=1$.

Foreground avoidance in the  form used here is reasonable in the context of a simple Fisher forecast approach. However, this approach does not incorporate the systematics that further complicate foreground removal, including for example polarisation leakage, radio frequency interference,  and beam effects. A complete treatment needs to include foreground removal and systematics in the data pipeline; see e.g. \cite{Spinelli:2021emp,Crichton:2021hlc,Fornazier:2021ini,Wang:2020lkn,Li:2020bcr,Matshawule:2020fjz,Liu:2019awk}
for recent work in this direction.

\section{Methodology}\label{method}
\subsection{Fisher matrix}\label{sec:fisher}

The Fisher matrix formalism is used to predict constraints on cosmological parameters and distance measures.  In a redshift bin at $z_i$, the Fisher matrices of the HI IM power spectrum is \cite{Tegmark1997,Seo_2003},
\begin{equation}\label{eq:fisherPs}
    F_{\alpha\beta}^{P}(z_i)= {1\over2} \sum_k\int_{-1}^1\!\! {\rm d}\mu \frac{\partial P_{\rm HI}^{\rm obs}(\bk,z_i)}{\partial \theta_{\alpha}}\frac{\partial P_{\rm HI}^{\rm obs}(\bk,z_i)}{\partial \theta_{\beta}}\frac{1}{\Delta P^2(\bk,z_i)} \, ,
   \end{equation}
while for the bispectrum 
   \begin{equation}\label{eq:fisherBs}
  F_{\alpha\beta}^{B}(z_i)=\frac{1}{4\pi}\!\sum_{k_1,k_2,k_3}\!\int_{-1}^1\!\! {\rm d}\mu_1 \!\! \int_0^{2\pi}\!\! {\rm d} \phi \frac{\partial B_{\rm HI}^{\rm obs}(\bk_1,\bk_2,\bk_3,z_i)}{\partial \theta_{\alpha}}\frac{\partial B_{\rm HI}^{\rm obs}(\bk_1,\bk_2,\bk_3,z_i)}{\partial \theta_{\beta}}\frac{1}{\Delta B^2(\bk_1,\bk_2,\bk_3,z_i)}.
  \end{equation}
Here $\theta_{\alpha}$ are the parameters to be constrained, the sum over triangles has $k_{\rm min}\le k_3\le k_2\le k_1 \le k_{\rm max}$, and $k_1$, $k_2$ and $k_3$ satisfy the triangle inequality. The bin size $\Delta k$ is taken to be the fundamental frequency of the survey, $k_{\rm f}=2\pi/L$, where for simplicity we approximate the survey volume as a cube, $L=V_{\rm survey}^{1/3}$. 
In the case of the bispectrum, we exploit the azimuthal symmetry of the RSD tree-level expression to change the $\phi$ integration limits to $[\pi/2$, $3\pi/2]$ and to multiply the integral  by a factor of 2, speeding up the numerical calculations significantly.

 \begin{figure*}[t]
\centering
\resizebox{0.5\textwidth}{!}{\includegraphics{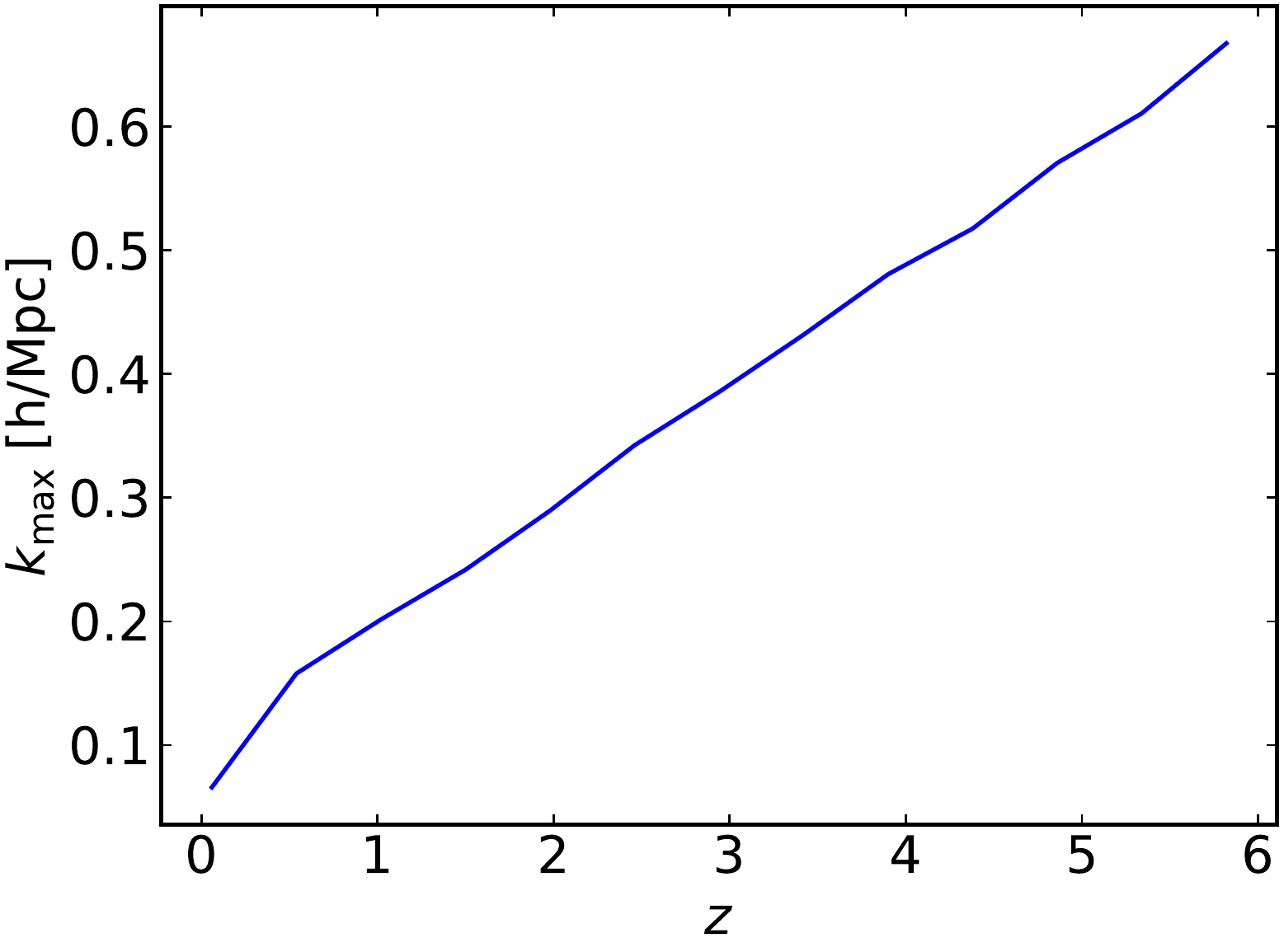}}
  \caption{Maximum Fourier mode corresponding to minimum scale for perturbative analysis.  }
  
\label{figkmax}
\end{figure*}
 
 The minimum value of the wavenumber is $k_{\rm min}=k_{\rm f}$, \ie the largest scale available to the survey, while the maximum value $k_{\rm max}$ corresponds to the smallest scale where the theoretical model is reliable (see 
 \Cref{figkmax}). We follow \cite{Karagiannis:2020dpq} and set 
 \begin{equation}\label{kmax}
 k_{\rm max}(z)=0.75\,k_{\rm NL}(z)\quad\mbox{where}\quad k_{\rm NL}(z)^{-2} = \frac{1}{6\pi^2} \int_0^\infty {\rm d}k\, P^{\rm L}_{\rm m}(k,z).    
 \end{equation}
Here $k_{\rm NL}$ is given by the one-dimensional velocity dispersion. The choice of $k_{\rm max}$ confines the analysis within the perturbative regime, where the tree-level description offers a good agreement with numerical results \cite{Gil-Marin:2014sta,Lazanu2015b,Hashimoto:2017klo,Chan2017,Oddo:2019run}.

 The full set of parameters, which encapsulate the main contributions of uncertainty in our model, consists of 5 cosmological parameters and 11 further parameters in each redshift bin $i$:
 \begin{align}\label{eq:params}
 \bm{\theta}(z_i)=&\Big\{\Omega_{\rm b},\Omega_{\rm c},h,A_{\rm s},n_{\rm s};   
  D_A(z_i),H(z_i),f(z_i),\nonumber \\
  & b_1(z_i),b_2(z_i),b_{s^2}(z_i),\sigma_P(z_i),\sigma_B(z_i),P_\veps(z_i),P_{\veps\veps_{\delta}}(z_i),B_{\veps}(z_i) \Big\}.
  \end{align}
  We assume redshift bins are independent, so that e.g.  $\partial P(\bk,z_i)/\partial \theta_\alpha(z_j)=0$. Then the total Fisher matrix is
  \begin{equation}\label{eq:totFISHER}
    F_{\alpha\beta}^{\rm tot}=\sum_{i=1}^{N_z}F_{\alpha\beta}(z_i). 
  \end{equation} 
Here $N_z$ is the number of redshift bins: HIRAX -- $N_z=17$; MeerKAT L, UHF bands -- $N_z=5,~8 $;  SKAO 1, 2 bands -- $N_z=26,~5$. The choices for $N_z$ listed here, result in redshift bins of width $\Delta z\simeq 0.1$ for all surveys. 
  For the redshift-independent cosmological parameters, the summation over all redshift bins is performed as in \Cref{eq:totFISHER}, leading to a $5\times 5$ block in the final $F_{\alpha\beta}^{\rm tot}$, which corresponds to the summed contribution from the total redshift range. Hence, this process will take the $5+11$ square Fisher matrix from each redshift and combine them all to form the total Fisher matrix, which includes all redshift bins and has dimension $5+11N_z$. Further details can be found in \cite{Euclid:2019clj}.

  The stochastic bias contributions  $P_\veps,P_{\veps\veps_{\delta}}$ and $B_\veps$, as well as the FoG parameters  $\sigma_P$ and $\sigma_B$, are considered nuisance parameters and are marginalised over. Once the final total Fisher matrix is constructed, its inverse yields the minimum error on a parameter as $\sigma(\theta_\alpha)=\sqrt{(F^{-1})_{\alpha\alpha}}$, in the case of the power spectrum and bispectrum. The forecasts from the summed power spectrum and bispectrum signal are also considered, by adding together the corresponding total Fisher matrices, \ie $F_{\alpha\beta}^{P+B}=F_{\alpha\beta}^{ P}+F_{\alpha\beta}^{ B}$. The cross-Fisher between the power spectrum and bispectrum is neglected, since its impact on the final constraints, for the parameters of interest, is minimal \cite{Yankelevich2018}.

  \subsection{Cosmological models}\label{sec:cosmo_models}
  
  \begin{table}[t]
 \centering
\begin{tabular}{||c c c c c c c c||} 
 \hline
 \multicolumn{5}{||c}{$\Lambda$CDM} & \multicolumn{3}{c||}{Extensions} \\ 
 \hline
 $\Omega_{\rm b}h^2$ & $\Omega_{\rm c}h^2$ & $h$ & $n_{\rm s}$ & $10^9A_{\rm s}$ & $w_0$ & $w_a$ & $\gamma$ \\ [0.5ex]
 \hline\hline
 $0.02237$ & $0.12$ & $0.6736$ & $0.9649$ & $2.1$ & $-1$ & $0$ & $0.55$ \\ [1ex]
 \hline
\end{tabular} 
\caption{Fiducial $\Lambda$CDM cosmological parameters, as measured by {\em Planck} \cite{Planck2018_cosmo}, and fiducial parameters for the extensions considered here.}
\label{table:fid_cosmo}
\end{table}
  
After marginalising over the nuisance parameters, the constraints on the remaining redshift-independent and redshift-dependent parameters, containing the key cosmological information, are transformed into constraints on parameters of the cosmological model. The baseline  model is the spatially flat $\Lambda$CDM model, defined by 5 parameters:
    \begin{equation}
\text{$\Lambda$CDM}:\big\{{\Omega_\text{b}},{\Omega_\text{c}}, h, n_\text{s}, A_{\rm s} \big\}  .
\label{eq:LCDM}
\end{equation}
To investigate further the constraining power of future HI IM surveys on deviations from the $\Lambda$CDM model, we consider two minimal extensions:
    \begin{itemize}
        \item modification of the equation of state of the dark energy:
\begin{equation}\label{eq:wde}
 w_{\rm de}(z)=w_0+w_a\, {z\over 1+z} \,;  
\end{equation}
        \item modification in the growth rate of large-scale structure, via the growth index $\gamma$ \cite{Lahav1991,Linder2005}:
\begin{equation}\label{eq:fgamma}
f(z)=\big[\Omega_{\rm m}(z)\big]^\gamma   \,, 
\end{equation}
where a significant deviation from $\gamma=0.55$ indicates either modified gravity or non-standard (e.g. clustering) dark energy.
    \end{itemize}
The extension models are defined by
  \begin{align}
\text{\wCDM}&:
\big\{{\Omega_\text{b}},{\Omega_\text{c}}, h, n_\text{s}, A_{\rm s}, w_0, w_a \big\},
\label{eq:wcdm}\\
\text{\gCDM}&:\big\{{\Omega_\text{b}},{\Omega_\text{c}}, h, n_\text{s}, A_{\rm s}, \gamma \big\}  .
\label{eq:gcdm}
\end{align}
Fiducial parameter values for all models are shown in \Cref{table:fid_cosmo}. 
  
The projection from the Fisher matrix of the initial parameter set $\theta_\alpha$ in \Cref{eq:params} (after marginalising the nuisance) into the parameters $\tilde\theta_A$ of the models in \Cref{eq:LCDM,eq:wcdm,,eq:gcdm}, is 
performed via the Jacobian transformation, 

\begin{equation}
      \widetilde{F}_{AB}
      =\big(J^{\rm T}\,F^{\rm tot}\, J \big)_{AB}
      =\sum_{\alpha,\beta}J_{A\alpha}\, F_{\alpha\beta}^{\rm tot}\,J_{\beta B} \qquad\mbox{where}\quad J_{A\alpha}={\partial \theta_\alpha \over \partial \tilde\theta_A} \,.
\label{eq:Jacobian}
\end{equation}
The precision of a particular survey on two specific parameters can be quantified through a figure of merit (FoM) \cite{Albrecht:2006um}, which is inversely proportional to the area of the $2\sigma$ contour of the two parameters, after marginalising over the rest. The `dark energy' FoM is usually taken as the FoM for $w_0$ and $w_a$:
\begin{equation}\label{eq:FoM}
    {\rm FoM}_{w_0w_a}=\left[ {\rm det}\left(\widetilde{F}_{w_0w_a}\right) \right]^{1/2}.
\end{equation}

\subsection{Statistical error} 
 
In a Gaussian approximation to the covariances of the power spectrum and  bispectrum,  the off-diagonal terms and the cross-covariance of $P$ and $B$ are neglected.
Then the variances for the two correlators are \cite{Sefusatti2006,Sefusatti2007}:
 \begin{align}
 \Delta P^2(\bk,z)&=\frac{4\pi^2}{V_{\text{survey}}(z)k^2\Delta k(z)}P_{\rm HI}(\bk,z)^2\, , \label{eq:deltaP2} \\
 \Delta B^2(\bk_1,\bk_2,\bk_3,z)&=s_{123}\,\pi\, k_{\rm f}(z)^3\,\frac{P_{\rm HI}(\bk_1,z)\,P_{\rm HI}(\bk_2,z)\,P_{\rm HI}(\bk_3,z)}{k_1k_2k_3\,[\Delta k(z)]^3}\, .\label{eq:deltaB2}
  \end{align}
Here $P_{\rm HI}$ includes the thermal noise and
$s_{123}=6,2,1$, for equilateral, isosceles and non-isosceles triangles respectively. In addition, for degenerate configurations, \ie $k_i=k_j+k_m$, the bispectrum variance should be multiplied by a factor of 2 \cite{Chan2017,Desjacques2016}. 
  
The Gaussian approximation is assumed to be accurate enough up to linear and mildly nonlinear scales and for high-density samples considered here, for both power spectrum and bispectrum, following \cite{Howlett:2017vwp,Barreira:2017kxd,Li:2018scc,Blot:2018oxk,Chan2017}. Off-diagonal terms of the covariances are related to higher-order loop corrections \cite{Sefusatti2006}, making the numerical implementation extremely tedious. They become important at small scales, where the Gaussian approximation breaks down. In the case of the bispectrum, these corrections affect even the variance and can have a significant effect on cosmological parameters \cite{Chan2017}. 

It has also been shown that non-Gaussian contributions in the form of off-diagonal terms in the bispectrum covariance and a non-zero cross-covariance, become important even on large scales for squeezed configurations, and
neglecting these contributions can lead to serious under-estimation of errors on local primordial non-Gaussianity \cite{Gualdi:2020ymf,Biagetti:2021tua,Floss:2022wkq}. We do not consider primordial non-Gaussianity here and we expect that the effect on our error estimates is smaller. Nevertheless, further work is needed in order to include non-Gaussian effects  for robust error estimates on all cosmological parameters.

As a first step in this direction, we include non-Gaussian corrections to the diagonal part of the bispectrum covariance, following  the prescription of \cite{Chan2017}:
  \begin{align}\label{eq:DB2_NL}
 \Delta B_\text{NL}^2(\bk_1,\bk_2,\bk_3)=& {}\Delta B^2(\bk_1,\bk_2,\bk_3) \nonumber \\
 &+\frac{s_{123}\,\pi\, k_{\rm f}^3}{k_1k_2k_3\,(\Delta k)^3}\Big[P_{\rm HI}(\bk_1)\,P_{\rm HI}\,(\bk_2)\,P_{\rm HI}^\text{NL}(\bk_3)+2\perm \Big],
\end{align}
where we  omit the $z$-dependence for brevity. The nonlinear power spectrum $P_{\rm HI}^\text{NL}(\bk)$ is given by \Cref{eq:Pgs} after the replacement: $P_{\rm m}(k)\rightarrow P^{\rm NL}_{\rm m}(k)-P_{\rm m}^{\rm L}(k)$, where $P^{\rm NL}_{\rm m}$ is the nonlinear matter power spectrum, described in \Cref{sec:RSDmodel}.

\subsection{Theoretical error}\label{sec:theoretical_er} 

 At small scales the statistical error becomes minimal, allowing for an increased signal and tighter constraints on cosmological and distance parameters, especially in the case of the bispectrum \cite{Hahn:2019zob,Hahn:2020lou}. However, on these scales the perturbative approach fails to describe the clustering of tracers. Even within the perturbative regime, as we approach  non-linear scales, higher-order loop corrections become important. Introducing a sharp $k_{\rm max}$ cut-off excludes all scales beyond the validity of the chosen model. Nonetheless, the importance of  loop corrections is gradual, indicating that neglecting them  introduces biases at any $k_{\rm max}$.  In the Fisher matrix analysis performed here, the uncertainty from excluding next-to-leading order corrections will be taken into account via the theoretical error approach introduced in \cite{Baldauf2016}.

 In this formalism, theoretical errors are defined as the difference between the chosen perturbative order (\eg, tree-level) and the next higher-order (\eg, 1-loop).  The theoretical error acts as correlated noise, forming the following covariance for the power spectrum:
 \begin{equation}
     C^E_{P}=E_P(\bk,z)E_P(\bk',z)\exp\left[-\frac{(k-k')^2}{2\,\Delta k^2}\right],
 \end{equation}
 while for the bispectrum
  \begin{equation}
     C^E_{ B}=E_B(\bk_1,\bk_2,\bk_3,z)E_B(\bk_1',\bk_2',\bk_3',z)\exp\left[-\sum_{i=1}^{3}\frac{(k_i-k_i')^2}{2\, \Delta k^2}\right].
 \end{equation}
  These theoretical error covariances are added to the statistical variances, presented in the previous section, to form the final covariances that are used in the Fisher matrix formalism.  The correlation length $\Delta k$ cannot be very small because the theoretical error covariance would be uncorrelated between the different momentum configurations. Here we use the value $\Delta k=0.05\;\Mpc$, proposed by \cite{Baldauf:2015aha}, which is motivated by the scales of the BAO wiggles. Note that the final theoretical error covariance is independent of the value of the correlation length, as long as the size of the $k$-bin is much smaller than $
  \Delta k$. This is true for the surveys we consider here, since the chosen binning size in the momentum space is the fundamental frequency, and  $k_f\ll\Delta k$.  
  
  The envelopes $E_P$ and $E_B$ are given from the fitting to the desired high-order correction. For the power spectrum we use the envelope, fitted to the explicit 2-loop calculations \cite{Baldauf2016}:
    \begin{equation}
        E_P(\bk,z)=D(z)^4P_{\rm HI}(\bk,z)\left(\frac{k}{0.45\;\Mpc}\right)^{3.3},
    \end{equation}
where $P_{\rm HI}$ is given by \Cref{eq:Pgs}, but without the thermal noise contribution. For the bispectrum, we use again an envelope  fitted against the 2-loop calculations \cite{Baldauf2016}: 
     \begin{equation}
        E_B(\bk_1,\bk_2,\bk_3,z)=3D(z)^4B_{\rm HI}(\bk_1,\bk_2,\bk_3,z)\left(\frac{k_T}{0.45\;\Mpc}\right)^{3.3},
    \end{equation}
where $B_{\rm HI}(\bk1,\bk_2,\bk_3,z)$ is given by \Cref{eq:Bgs} and $k_T=(k_1+k_2+k_3)/3$. 

Note that the theoretical error approach, briefly described here, does not take into account FoG effect, which is not captured by perturbation theory and can become important at the loop correction level (see \cite{Chudaykin:2019ock} for a discussion). Nonetheless, due to the scales considered here, the exclusion of FoG effects from the theoretical error approach is not expected to affect significantly the error covariances. 
    
The analysis here is mostly confined to scales where the tree-level description gives accurate predictions. Hence, we do not expect the theoretical uncertainties to significantly affect our forecasts. Nonetheless, the inclusion of theoretical errors is done in order to have, as much as possible, a complete characterisation of the covariance included in the Fisher formalism.

\subsection{Priors}\label{sec:priors}
 
The forecast results produced in this work, come from the Fisher information matrices. This is equivalent to assuming that all the information on the parameters comes from the likelihood, while very diffused priors are adopted. Therefore, the Fisher matrix results will be combined with the information on cosmological parameters coming from the observation of CMB performed by \Planck \cite{Planck2018_cosmo}. In order to do this, we use the Markov chain that samples the posterior, from the \Planck webpage\footnote{\url{http://pla.esac.esa.int/pla/\#cosmology}}, which corresponds to each fiducial cosmological model considered (\Cref{sec:cosmo_models}). From the chains we compute the covariance matrix that corresponds to the subset of parameters considered here and proceed to invert the matrix in order to get the \Planck Fisher matrix. The latter is then summed to the Fisher matrices of the power power spectrum and bispectrum, as well as their joined case. Effectively, we treat the \Planck likelihood as a multivariate Gaussian, which is sufficient for the free cosmological parameters considered here.

\section{Results}\label{sec:results}


In the previous sections we outline the model, the surveys and methodology used to produce our forecasts. In this section we summarize the results on estimating the constraints on cosmological parameters coming from next-generation HI intensity experiments, in the case of the $\Lambda$CDM, \gCDM and \wCDM models, as well as on redshift dependent quantities, like cosmic distances and bias parameters. For all parameters, the forecasts are derived from the HI power spectrum, bispectrum and their joint signal (see \Cref{method} for details), while \Planck priors are considered throughout, unless otherwise stated.

\subsection{Cosmological parameters}\label{sec:cosmo_res}

\begin{figure*}[t]
\centering
\resizebox{\textwidth}{!}{\includegraphics{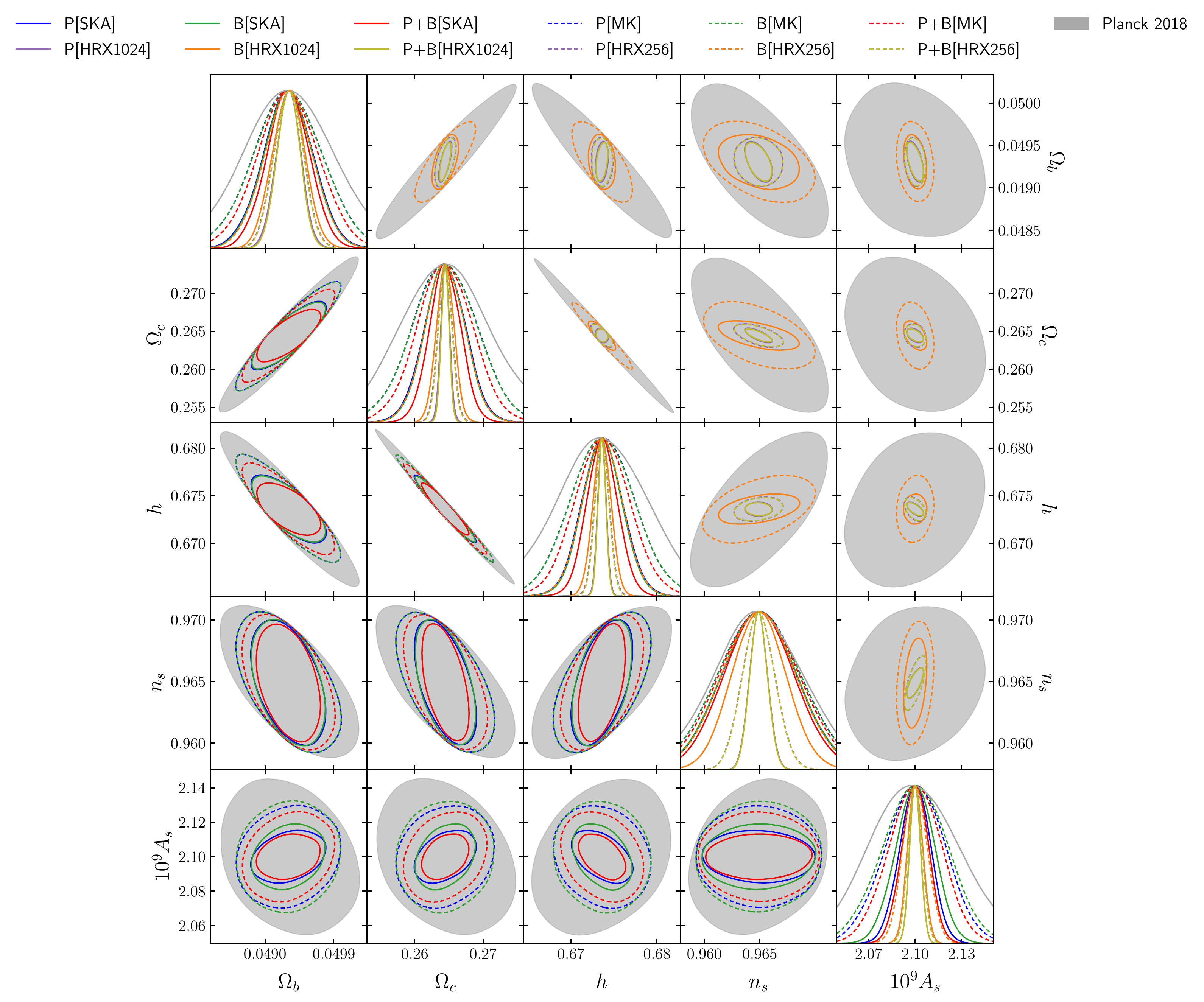}}
  \caption{Marginalised 2D contours of 1$\sigma$ errors for  $\Lambda$CDM parameters. Panels below the diagonal give SD surveys, panels above give IF surveys. Colour and line-type are shown at the top. All results include \Planck priors; \Planck results are presented for comparison. Contours correspond to the  foreground cut, $\kparmin=0.01\;\Mpc$.}  
  
\label{fig:LCDM_triangle_plots}
\end{figure*}

Here we present our results on  the standard $\Lambda$CDM cosmological model, by combining the \Planck measurements with the forecasts coming from HI intensity mapping surveys in single-dish mode (MeerKAT and SKAO) and in interferometer mode (HIRAX-256 and HIRAX-1024). In the case of the SD mode experiments which operate in two frequency bands, the combined signal from both bands is considered (see \Cref{table:survey_specs_SD}). We exclude the redshift bins within the range  $0.4\leq z \leq 0.58$ (MeerKAT UHF band)  and $0.35\leq z\leq 0.49$ (SKAO band 1), in order to avoid double-counting the information within the overlapping redshifts. The marginalised relative errors on each  $\Lambda$CDM cosmological parameter are presented in \Cref{table:results_SD} (SD mode) and \Cref{table:results_IF} (IF mode), while the 2D contours of the forecasts are shown in \Cref{fig:LCDM_triangle_plots}.

 The MeerKAT survey, once combined with \Planck, provides percent-level constraints on all cosmological parameters, where the HI power spectrum and bispectrum give very similar relative errors (see \Cref{table:results_SD}). Due to the CMB external information, the summed signal from the two- and three-point correlators offers only a small improvement. As indicated by the panels in \Cref{fig:LCDM_triangle_plots}, MeerKAT provides a marginal advantage over the \Planck constraints (grey shaded contour) for most of the cosmological parameters, where the power spectrum constraints are only slightly better than those by the bispectrum. 
 
SKAO produces moderately improved results compared to MeerKAT for all parameters, due to the larger volume and redshift range probed by SKAO, which increases the Fourier space resolution, thus improving the overall signal of the summary statistics considered. This is more evident for $\Omega_{\rm c}$ and $A_{\rm s}$. In the case of the latter,  SKAO reduces the \Planck errors two-fold (see bottom panels of \Cref{fig:LCDM_triangle_plots}).  Similarly to MeerKAT, the HI power spectrum and bispectrum of SKAO yield approximately equal relative errors, and the summed signal of the two correlators produces a more notable improvement compared to MeerKAT.

\begin{table}[t]
\centerline{
\begin{tabular} { l c c c c c c }
\noalign{\vskip 3pt}\hline\noalign{\vskip 1.5pt}\hline\noalign{\vskip 6pt}
 \multicolumn{1}{c}{\bf } & \multicolumn{3}{c}{\bf MeerKAT} & \multicolumn{3}{c}{\bf SKAO} \\[7pt]
 $[\%]$ & \bf P & \bf B & \bf P+B & \bf P & \bf B & \bf P+B \\ 
\noalign{\vskip 3pt}\hline\noalign{\vskip 6pt}
\multicolumn{7}{c}{$\Lambda$CDM}  \\
\noalign{\vskip 3pt}\hline\noalign{\vskip 6pt}
$\sigma(\Omega_{\rm b})/\Omega_{\rm b}$ &  0.92 (0.87) & 0.92 (0.84) & 0.81 (0.75) & 0.66 (0.61) & 0.64 (0.58) & 0.56 (0.52)  \\  

$\sigma(\Omega_{\rm c})/\Omega_{\rm c}$ &  1.81 (1.68) & 1.8 (1.62) & 1.54 (1.37) & 1.13 (1.0) & 1.1 (0.92) & 0.86 (0.74)  \\  

$\sigma(h)/h$ &  0.56 (0.52) & 0.56 (0.5) & 0.47 (0.42) & 0.35 (0.31) & 0.34 (0.29) & 0.27 (0.23)  \\  

$\sigma(n_{\rm s})/n_{\rm s}$ &  0.39 (0.38) & 0.39 (0.38) & 0.37 (0.37) & 0.35 (0.34) & 0.35 (0.34) & 0.33 (0.32)  \\  

$\sigma(10^{9}A_{\rm s})/10^{9}A_{\rm s}$ &  0.93 (0.92) & 1.01 (0.95) & 0.82 (0.78) & 0.48 (0.47) & 0.6 (0.54) & 0.42 (0.39)  \\    
  
\noalign{\vskip 3pt}\hline\noalign{\vskip 6pt}
\multicolumn{7}{c}{\gCDM}  \\
\noalign{\vskip 3pt}\hline\noalign{\vskip 6pt}
$\sigma(\Omega_{\rm b})/\Omega_{\rm b}$ &  0.95 (0.88) & 0.95 (0.85) & 0.83 (0.75) & 0.66 (0.61) & 0.65 (0.58) & 0.56 (0.52)  \\  

$\sigma(\Omega_{\rm c})/\Omega_{\rm c}$ &  1.88 (1.7) & 1.9 (1.65) & 1.59 (1.37) & 1.14 (1.0) & 1.12 (0.92) & 0.87 (0.74)  \\  

$\sigma(h)/h$ &  0.58 (0.53) & 0.59 (0.51) & 0.49 (0.43) & 0.35 (0.31) & 0.35 (0.29) & 0.27 (0.23)  \\  

$\sigma(n_{\rm s})/n_{\rm s}$ &  0.39 (0.38) & 0.39 (0.38) & 0.37 (0.37) & 0.35 (0.34) & 0.35 (0.34) & 0.33 (0.32)  \\  

$\sigma(10^{9}A_{\rm s})/10^{9}A_{\rm s}$ &  1.22 (1.17) & 1.26 (1.17) & 1.13 (1.06) & 0.65 (0.63) & 0.88 (0.77) & 0.58 (0.54)  \\  

$\sigma(\gamma)/\gamma$ &  12.67 (11.8) & 14.89 (12.69) & 10.73 (9.45) & 6.11 (5.7) & 8.11 (6.57) & 5.05 (4.45)  \\

\noalign{\vskip 3pt}\hline\noalign{\vskip 6pt}
\multicolumn{7}{c}{\wCDM}  \\
\noalign{\vskip 3pt}\hline\noalign{\vskip 6pt}

$\sigma(\Omega_{\rm b})/\Omega_{\rm b}$ &  5.06 (4.86) & 5.01 (4.5) & 3.95 (3.6) & 2.48 (2.34) & 2.55 (2.21) & 1.86 (1.68)  \\  

$\sigma(\Omega_{\rm c})/\Omega_{\rm c}$ &  5.11 (4.88) & 4.99 (4.44) & 3.93 (3.54) & 2.46 (2.27) & 2.49 (2.11) & 1.78 (1.55)  \\  

$\sigma(h)/h$ &  2.26 (2.17) & 2.23 (2.0) & 1.75 (1.59) & 1.07 (1.01) & 1.11 (0.95) & 0.78 (0.69)  \\  

$\sigma(n_{\rm s})/n_{\rm s}$ &  0.42 (0.42) & 0.42 (0.42) & 0.42 (0.41) & 0.4 (0.4) & 0.4 (0.4) & 0.39 (0.38)  \\  

$\sigma(10^{9}A_{\rm s})/10^{9}A_{\rm s}$ &  1.44 (1.42) & 1.42 (1.38) & 1.33 (1.29) & 1.05 (1.04) & 1.1 (1.06) & 0.93 (0.9)  \\  

$\sigma(w_0)/w_0$ &  31.63 (30.29) & 35.04 (31.43) & 25.41 (23.3) & 13.69 (12.78) & 16.42 (14.13) & 10.41 (9.3)  \\  

$\sigma(w_\alpha)/w_\alpha$ &  103.46 (99.03) & 130.64 (117.1) & 86.18 (79.42) & 40.56 (37.75) & 55.86 (48.32) & 32.35 (29.06)  \\  
\noalign{\vskip 3pt}\hline\noalign{\vskip 0.5pt}\hline\noalign{\vskip 6pt}

$\rm FoM$ & 10.8 (11.9) & 7.8 (9.8) & 17.3 (20.7) & 73 (82) & 44.7 (59) & 120.5 (147) \\

$\rm FoM_{\rm nPp} $ & 5.5 (6.2) & 5.5 (7.3) & 11.9 (14.8) & 52.7 (60.4) & 39 (52.7) & 99.5 (123)\\

\noalign{\vskip 3pt}\hline\noalign{\vskip 1.5pt}\hline\noalign{\vskip 5pt}
\end{tabular}

}

    \caption{Forecasts of marginalised 1$\sigma$ relative errors (in \%) on cosmological parameters in $\Lambda$CDM, \gCDM and \wCDM, for  SD surveys MeerKAT and SKAO, using the combined signal of both bands of each survey (see \Cref{table:survey_specs_SD}), avoiding double-counting in overlapping redshift bins. Columns display constraints from  power spectrum ({\bf P}), bispectrum ({\bf B}) and  combination ({\bf P+B}). Main results correspond to the  $\kparmin=0.01\;\Mpc$ foreground cut; the idealised case ($\kparmin=0$) is in  parenthesis. All results assume \Planck priors (see \Cref{sec:priors}). Last row shows the FoM (see \Cref{sec:cosmo_models}) for dark energy parameters $w_0$, $w_a$. FoM$_{\rm nPp}$ is the FoM  without \Planck priors. 
    }
    \label{table:results_SD}
\end{table}

\begin{table}[t]
\centerline{
\begin{tabular} { l c c c c c c }
\noalign{\vskip 3pt}\hline\noalign{\vskip 1.5pt}\hline\noalign{\vskip 6pt}
 \multicolumn{1}{c}{\bf } & \multicolumn{3}{c}{\bf HIRAX-256} & \multicolumn{3}{c}{\bf HIRAX-1024} \\[7pt]
 $[\%]$ & \bf P & \bf B & \bf P+B & \bf P & \bf B & \bf P+B \\ 
\noalign{\vskip 3pt}\hline\noalign{\vskip 6pt}
\multicolumn{7}{c}{$\Lambda$CDM}  \\
\noalign{\vskip 3pt}\hline\noalign{\vskip 6pt}
$\sigma(\Omega_{\rm b})/\Omega_{\rm b}$ &  0.39 (0.37) & 0.64 (0.61) & 0.38 (0.36) & 0.32 (0.32) & 0.43 (0.42) & 0.31 (0.3)  \\  

$\sigma(\Omega_{\rm c})/\Omega_{\rm c}$ &  0.39 (0.23) & 1.11 (1.03) & 0.37 (0.23) & 0.24 (0.22) & 0.48 (0.46) & 0.22 (0.2)  \\  

$\sigma(h)/h$ &  0.12 (0.08) & 0.35 (0.32) & 0.12 (0.08) & 0.07 (0.07) & 0.16 (0.15) & 0.07 (0.07)  \\  

$\sigma(n_{\rm s})/n_{\rm s}$ &  0.15 (0.15) & 0.34 (0.34) & 0.15 (0.14) & 0.08 (0.08) & 0.25 (0.24) & 0.08 (0.08)  \\  

$\sigma(10^{9}A_{\rm s})/10^{9}A_{\rm s}$ &  0.22 (0.22) & 0.38 (0.36) & 0.22 (0.22) & 0.17 (0.17) & 0.24 (0.23) & 0.17 (0.17)  \\  

\noalign{\vskip 3pt}\hline\noalign{\vskip 6pt}
\multicolumn{7}{c}{\gCDM}  \\
\noalign{\vskip 3pt}\hline\noalign{\vskip 6pt}
$\sigma(\Omega_{\rm b})/\Omega_{\rm b}$ &  0.39 (0.37) & 0.64 (0.61) & 0.39 (0.36) & 0.32 (0.32) & 0.43 (0.42) & 0.31 (0.31)  \\  

$\sigma(\Omega_{\rm c})/\Omega_{\rm c}$ &  0.4 (0.25) & 1.12 (1.03) & 0.38 (0.25) & 0.29 (0.23) & 0.49 (0.46) & 0.25 (0.21)  \\  

$\sigma(h)/h$ &  0.13 (0.09) & 0.35 (0.32) & 0.12 (0.09) & 0.1 (0.08) & 0.16 (0.15) & 0.09 (0.08)  \\  

$\sigma(n_{\rm s})/n_{\rm s}$ &  0.17 (0.15) & 0.34 (0.34) & 0.16 (0.15) & 0.1 (0.09) & 0.25 (0.24) & 0.09 (0.09)  \\  

$\sigma(10^{9}A_{\rm s})/10^{9}A_{\rm s}$ &  0.24 (0.24) & 0.56 (0.5) & 0.24 (0.23) & 0.17 (0.17) & 0.28 (0.27) & 0.17 (0.17)  \\  

$\sigma(\gamma)/\gamma$ &  2.41 (2.04) & 16.3 (14.62) & 2.34 (1.93) & 1.21 (1.0) & 7.51 (6.84) & 1.13 (0.97)  \\  

\noalign{\vskip 3pt}\hline\noalign{\vskip 6pt}
\multicolumn{7}{c}{\wCDM}  \\
\noalign{\vskip 3pt}\hline\noalign{\vskip 6pt}

$\sigma(\Omega_{\rm b})/\Omega_{\rm b}$ &  1.04 (0.96) & 3.61 (3.38) & 1.01 (0.93) & 0.74 (0.69) & 1.81 (1.69) & 0.68 (0.64)  \\  

$\sigma(\Omega_{\rm c})/\Omega_{\rm c}$ &  0.95 (0.79) & 3.53 (3.3) & 0.92 (0.77) & 0.75 (0.66) & 1.79 (1.66) & 0.69 (0.61)  \\  

$\sigma(h)/h$ &  0.43 (0.38) & 1.61 (1.51) & 0.42 (0.37) & 0.31 (0.29) & 0.8 (0.75) & 0.29 (0.27)  \\  

$\sigma(n_{\rm s})/n_{\rm s}$ &  0.25 (0.23) & 0.4 (0.39) & 0.24 (0.22) & 0.2 (0.18) & 0.33 (0.33) & 0.19 (0.17)  \\  

$\sigma(10^{9}A_{\rm s})/10^{9}A_{\rm s}$ &  0.45 (0.45) & 0.99 (0.96) & 0.45 (0.44) & 0.34 (0.3) & 0.64 (0.61) & 0.33 (0.29)  \\  

$\sigma(w_0)/w_0$ &  4.65 (4.1) & 18.36 (17.08) & 4.5 (3.99) & 3.13 (2.86) & 8.67 (8.03) & 2.88 (2.64)  \\  

$\sigma(w_\alpha)/w_\alpha$ &  12.6 (10.86) & 46.79 (43.68) & 12.14 (10.53) & 6.91 (6.32) & 20.74 (19.32) & 6.39 (5.85)  \\  

\noalign{\vskip 3pt}\hline\noalign{\vskip 0.5pt}\hline\noalign{\vskip 6pt}

$\rm FoM$ & 904 (1091) & 56 (62) & 993 (1209) & 3033 (3566) & 298 (334) & 3494 (4155) \\

$\rm FoM_{\rm nPp}$ & 775 (990) & 33 (39) & 873 (1119) & 2712 (3276) & 194 (226) & 3134 (3848)\\

\noalign{\vskip 3pt}\hline\noalign{\vskip 1.5pt}\hline\noalign{\vskip 5pt}
\end{tabular}

}

    \caption{As in \Cref{table:results_SD} but for the IF surveys considered here.}
    \label{table:results_IF}
\end{table}

SD mode is capable of probing better the large clustering scales \cite{Bull2015}, rendering a power spectrum analysis more appropriate. Nonetheless, the effect of the beam and the survey specifications allows for similar constraints from both correlators, once the \Planck measurements are considered. From the marginalised 2D contours in \Cref{fig:LCDM_triangle_plots}, we see that the bispectrum and power spectrum have almost the same correlations between parameters, due to the domination of the CMB signal. Adding the two- and three-point statistics offers a moderate improvement on all cosmological parameters, besides the spectral index, whose forecasts are seemingly unaffected by the inclusion of the bispectrum. This is evident for both SD surveys, as shown in \Cref{table:results_SD}. 

The power spectrum is known to suffer from degeneracies between cosmological and bias parameters. Adding the information from the bispectrum introduces new shape dependencies that break various parameter degeneracies, improving the overall forecasts on cosmological parameters. For instance, the bispectrum helps to break the notorious $A_{\rm s}$ and $b_1$ degeneracy present in the power spectrum, since its amplitude scales like $A_{\rm s}^2b_1^3$. In addition, the degeneracy between $A_{\rm s}$ and $f$ is broken in a similar way. The importance of adding the bispectrum to the power spectrum data has been shown previously \cite{GilMarin2014,GilMarin2017,Karagiannis2018,Yankelevich2018,Chudaykin:2019ock,Agarwal:2020lov,Ivanov:2021kcd}. Despite the initial expectations of a minimal contribution to the cosmological constraints, when SD surveys are considered, the HI bispectrum of MeerKAT and SKAO is capable of breaking or limiting the various degeneracies between parameters and thereby provides the improvement seen in the joint forecasts. This  is in agreement with recent findings for optical surveys \cite{Agarwal:2020lov,Ivanov:2021kcd}. 

The importance of adding the bispectrum can be also seen in cosmological models with a larger parameter space, since there are more degeneracies to break. This is true for the \gCDM and \wCDM models, where the addition of the bispectrum provides a significant improvement on all cosmological parameters, with the exception of $n_{\rm s}$. Naturally, this is more evident in the case of the model with the most free-parameters (\ie \wCDM), where the bispectrum not only provides an important contribution but actually dominates the joint forecasts for $\Omega_{\rm b}$, $\Omega_{\rm c}$ and $H_0$.

In the case of HIRAX, we consider the early-phase 256-element array and the future planned  1024-element array. For both setups, combining the HIRAX Fisher matrices  with the \Planck likelihoods, the errors on all $\Lambda$CDM parameters improve significantly, achieving sub-percent level, as shown by the contours of the upper triangular panels in \Cref{fig:LCDM_triangle_plots}. The constraints are mainly driven by the HI power spectrum, as is evident from the joint forecasts in \Cref{table:results_IF}, rendering the bispectrum contribution complementary. The combined signal from the two correlators for HIRAX-1024 provides the tightest constraints presented in this work. In particular, the joint power + bispectrum without any prior information, has a precision comparable to the recent \Planck results. 

The improvement in the forecasts between the initial and final arrays is attributed to the higher number of elements  and the amplitude of the baseline distribution $n_{\rm b}$, as given in \Cref{app:baseline}. This leads to a decrease in the instrumental noise [\Cref{eq:Pnoise_IF}] and therefore an increase in the range of available modes for the two- and three-point statistics.

 In IF surveys  the bispectrum has an advantage over the power spectrum, since IF mode can probe large to intermediate scales, forming a notable number of triangles and pushing the available modes to higher $k$ values, where the bispectrum signal is significantly boosted. This has been found to be true for parameters like the amplitude of primordial non-Gaussianity and bias in \cite{Karagiannis:2020dpq}. Despite this, for the cosmological parameters considered here, the full potential of the bispectrum is reduced in HIRAX. This is more evident in the high-noise case of HIRAX-256, where the bispectrum has negligible contribution to the final summed forecasts. HIRAX-1024,  keeping all  other survey specifications fixed, improves the bispectrum constraints more  than it does the power spectrum, reducing the difference between the forecasts from the two correlators for all cosmological parameters. 
 
 Contrary to what was argued in the SD case, the other cosmological models do not benefit appreciably from the bispectrum information. In the $\Lambda$CDM case, bispectrum constraints could be completely neglected. The reason lies in the instrumental noise of HIRAX, which damps the signal from relevant triangle configurations, reducing the potential of the bispectrum to break degeneracies and its contribution to constraints. 
 
 The HIRAX survey characteristics offer an important advantage to the power spectrum, within the considered scale range, which  provides sub-percent precision for all cosmological parameters, significantly improving over current CMB limits. Even without appreciable contribution from the bispectrum, the HIRAX forecasts are the tightest, significantly better than SKAO. In particular, the HIRAX-1024 power spectrum constraints are far stronger than those from SKAO power spectrum + bispectrum. This is despite the fact that SKAO spans a wider sky area and greater redshift range. The reason is SKAO's low dish density (which increases the instrumental noise) and strong beam effects (which reduces the range of $k$ modes on the most important scales). 
 
 An overall conclusion is that the synergy between the LSS and CMB data is crucial in achieving robust measurements on cosmological parameters from future surveys.

 \begin{figure*}[t]
\centering
\resizebox{\textwidth}{!}{\includegraphics{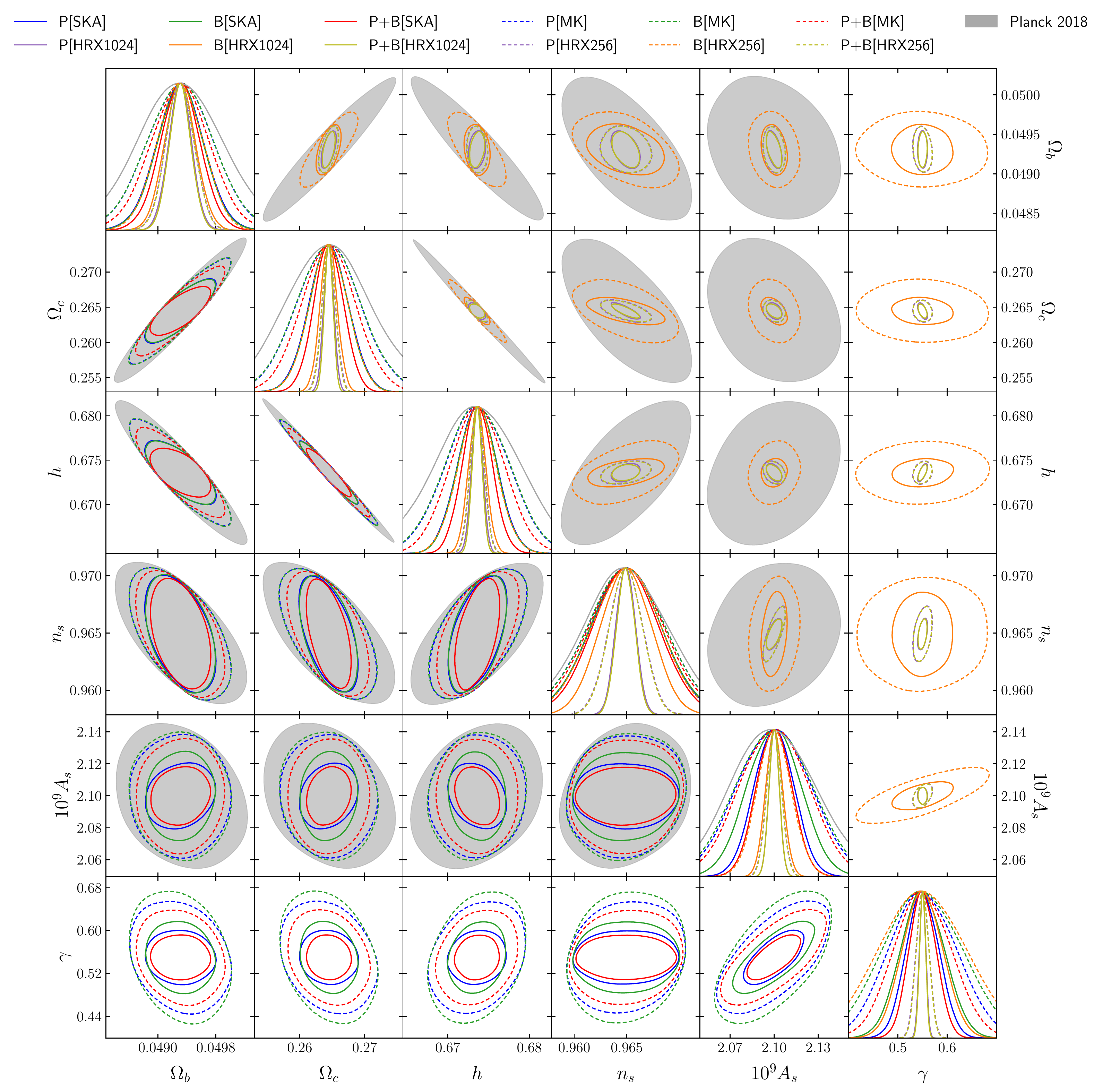}}
  \caption{As in \Cref{fig:LCDM_triangle_plots} but for the \gCDM model.}
  
\label{fig:gCDM_triangle_plots}
\end{figure*}

\subsection{Modified gravity}\label{sec:gamma}

Beyond the $\Lambda$CDM framework, the background and the perturbations can be modified. Modification of general relativity (GR) has been proposed as an alternative source for the accelerating expansion of the Universe observed at low redshifts \cite{Planck:2015bue,Joyce:2016vqv,Amendola:2016saw,Slosar:2019flp,Frusciante:2019xia}. There is a rich phenomenology for such models. We choose the simplest and extensively used way to probe modified growth of perturbations  by testing for deviations in the growth index $\gamma$ (see \Cref{eq:fgamma}) from its GR value $\gamma=0.55$ (which applies to
$\Lambda$CDM and also dynamical dark energy models that are non-clustering and non-interacting).

The bottom row of \Cref{fig:gCDM_triangle_plots}  presents the marginalised 2-D contours for $\gamma$ and the other cosmological parameters for SD surveys, while the last column is for IF surveys. Relative errors on $\gamma$  are presented in \Cref{table:results_SD} (SD surveys) and \Cref{table:results_IF}  (IF surveys). Note that the CMB data serve as priors only on the $\Lambda$CDM parameters (see \Cref{sec:cosmo_models}), in order to have a clear understanding on the capabilities of each survey to robustly constrain modified gravity models.

In the case of SD experiments, the growth index can be constrained with the modest precision of $\sim10-14\%$ by MeerKAT and $\sim5-8\%$ by SKAO. The power spectrum yields tighter constraints than the bispectrum, in agreement with the behaviour observed for the remaining cosmological parameters, discussed in \Cref{sec:cosmo_res}. Adding the bispectrum improves the power spectrum results for the two surveys by a few percent, as indicated by the joint forecasts shown in \Cref{table:results_SD}. SKAO shrinks the errors on $\gamma$  by a factor of 2 relative to MeerKAT. 

In the case of HIRAX, both power and bispectrum deliver a few-percent precision on $\gamma$, with the power spectrum being the main contributor, while the bispectrum has a complementary role, with forecasts that are almost 8 times less stringent. HIRAX-1024, combined with \Planck, will be capable of reaching a $1\%$ precision on $\gamma$, which would be important for eludicating the nature of gravity and any possible deviations from GR.

The marginalised contours of $\gamma$ in \Cref{fig:gCDM_triangle_plots} indicate that the inclusion of the growth index in the final parameter space has a minimal impact on the $\Lambda$CDM parameters. This is true for both correlators, whose contours seem to follow the same behaviour, and for all parameters, except for the amplitude of the primordial power spectrum, which exhibits sizeable degeneracies with $\gamma$. This is more evident in the case of the SD surveys, where for SKAO the errors on $A_{\rm s}$ increase by $\sim30\%$, as shown by the relative errors of the power spectrum and bispectrum presented in \Cref{table:results_SD}. For HIRAX, the $\gamma-A_{\rm s}$ degeneracy has a small effect on the constraints of both parameters.

\subsection{Dark energy}\label{sec:FOM}

Current observations are consistent with a cosmological constant $\Lambda$, a non-dynamical dark energy model with a constant equation of state, $w_\Lambda=-1$. In this section, we assess the sensitivity of future and current HI experiments to departures from a cosmological constant, by using a dynamical dark energy  fluid described by the redshift-dependent equation of state given in \Cref{eq:wde} (and with sound speed of 1, which ensures that the dark energy does not cluster).

\begin{figure*}[t]
\centering
\resizebox{\textwidth}{!}{\includegraphics{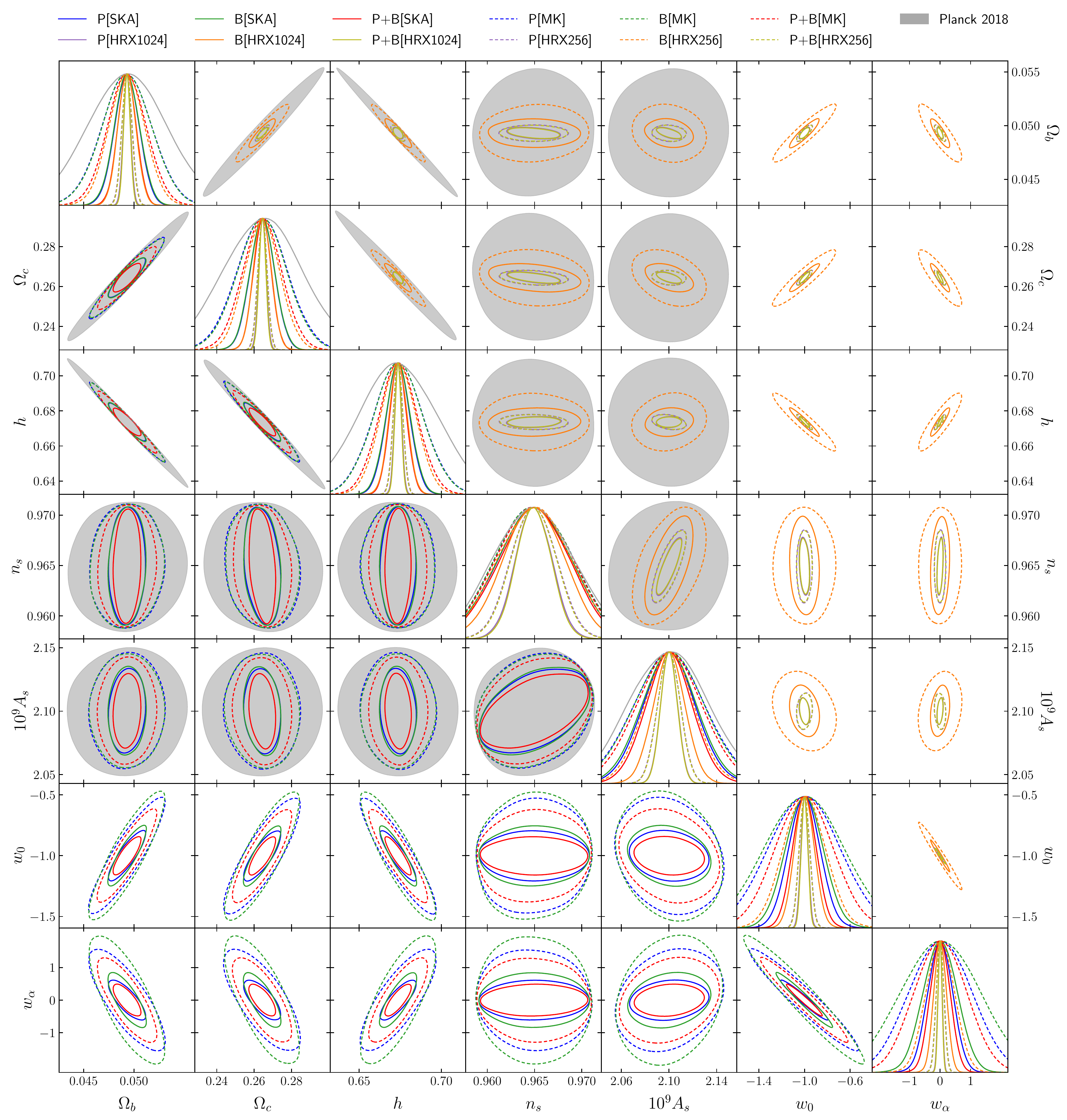}}
  \caption{As in \Cref{fig:LCDM_triangle_plots} but for the \wCDM model.}
  
\label{fig:wCDM_triangle_plots}
\end{figure*}

\Cref{fig:wCDM_triangle_plots}  presents the marginalised forecasts on \wCDM, where the contours on $w_0,w_a$ are in the two bottom rows and two right-most columns, for the case of SD and IF surveys respectively. The relative errors on $w_0$ and $w_a$, together with the other  cosmological parameters, are shown in  \Cref{table:results_SD,,table:results_IF}. 

In the case of SD surveys, power spectrum and bispectrum provide comparable constraints on $w_0$, where power-spectrum errors are a few percent smaller. Overall, MeerKAT achieves a moderate $25\%$ combined precision on $w_0$, while SKAO delivers  $\sim10\%$. This is not true for $w_a$, due to the substantial degeneracies between $w_0$ and $w_a$, as well as between the dark energy parameters and $\Omega_{\rm b}$, $\Omega_{\rm c}$ and $H_0$, as evident in the marginalised contours of \Cref{fig:wCDM_triangle_plots}. MeerKAT has no constraining power on $w_a$, while SKAO achieves $\sim30\%$ precision. 

On the other hand, both versions of HIRAX  achieve a significant improvement over the SD surveys. The constraints on $w_0$ and $w_a$ originate solely from the power spectrum, with negligible  contribution from the bispectrum. Although degeneracies between $w_0,w_a$ and cosmology remain (see upper triangle panels of \Cref{fig:wCDM_triangle_plots}), HIRAX provides enough signal to produce the most stringent constraints in this work,  reaching a few percent precision, when combining both correlators, on both parameters in the 1024-array setup. 

In order to assess the potential of surveys in constraining  dynamical dark energy,  we use the FoM in \Cref{eq:FoM}. We take the initial Fisher matrix with free parameters given by \Cref{eq:params} for all redshift bins (\ie $5+11N_z$ parameters), marginalise over the nuisance parameters and project the derived Fisher (with $5+6N_z$ parameters) to the  parameter space of \Cref{eq:wcdm}, and then marginalise out all parameters except $w_0$ and $w_a$. The subsequent FoM$_{w_0,w_a}$ results are reported in the bottom rows of  \Cref{table:results_SD,,table:results_IF}. These results include the CMB measurements, which are accounted for as priors on the $\Lambda$CDM parameters, while the FoM results, without the \Planck priors, are also presented under the label ${\rm FoM_{nPp}}$. 

For MeerKAT, the power spectrum and bispectrum have a similar contribution, where the addition of the bispectrum improves FoM by a factor of two. A similar improvement  from the inclusion of the bispectrum is observed for SKAO, but in this case the FoM from the power spectrum surpasses that from the bispectrum by $\sim35\%$. These findings indicate a partial breaking of degeneracies between the linear bias and primordial scalar amplitude and growth rate, when bispectrum measurements are included -- which then improves the constraining power on $w_0,w_a$. Adding  CMB information improves the FoM further, since $A_{\rm s}$ is very well determined by \Planck. The  enhancement for MeerKAT is $\sim50\%$,  for SKAO it is $\sim25\%$. 

In the case of HIRAX, the power spectrum provides the largest FoM, with the bispectrum contributing  $\sim10\%$, highlighting again the inferior role of the bispectrum in HIRAX forecasts. Once CMB data are added, the FoM from the HIRAX-256 power spectrum and combined correlators improves by $\sim15\%$, while from the bispectrum by a factor of $1.7$. For HIRAX-1024, the improvement on FoM  is $\sim12\%$ for the power spectrum and combined correlators, while for the bispectrum it is by a factor of $\sim1.5$. These findings indicate that  the bispectrum of HIRAX on its own is inadequate to break degeneracies between the considered parameters and therefore CMB data are necessary to improve the FoM, via the constraining power of \Planck on $A_{\rm s}$. The improvement provided by \Planck  in the HIRAX FoM is more significant for the correlator with the poorest signal, \ie the bispectrum. Combining the signal from both correlators for HIRAX-1024 with the CMB measurements, yields the highest FoM of this work, exceeding by far the FoM of \Planck and that of  recent LSS surveys (e.g. \cite{Yankelevich2018}). The HIRAX-1024  bispectrum alone surpasses by a significant amount the FoM from the SKAO combined correlators, in the case of the SD experiments (see  \Cref{table:results_SD,,table:results_IF}).

\subsection{Distance and growth rate measurements}\label{sec:distance}

\begin{figure*}[t]
\centering
\resizebox{\textwidth}{!}{\includegraphics{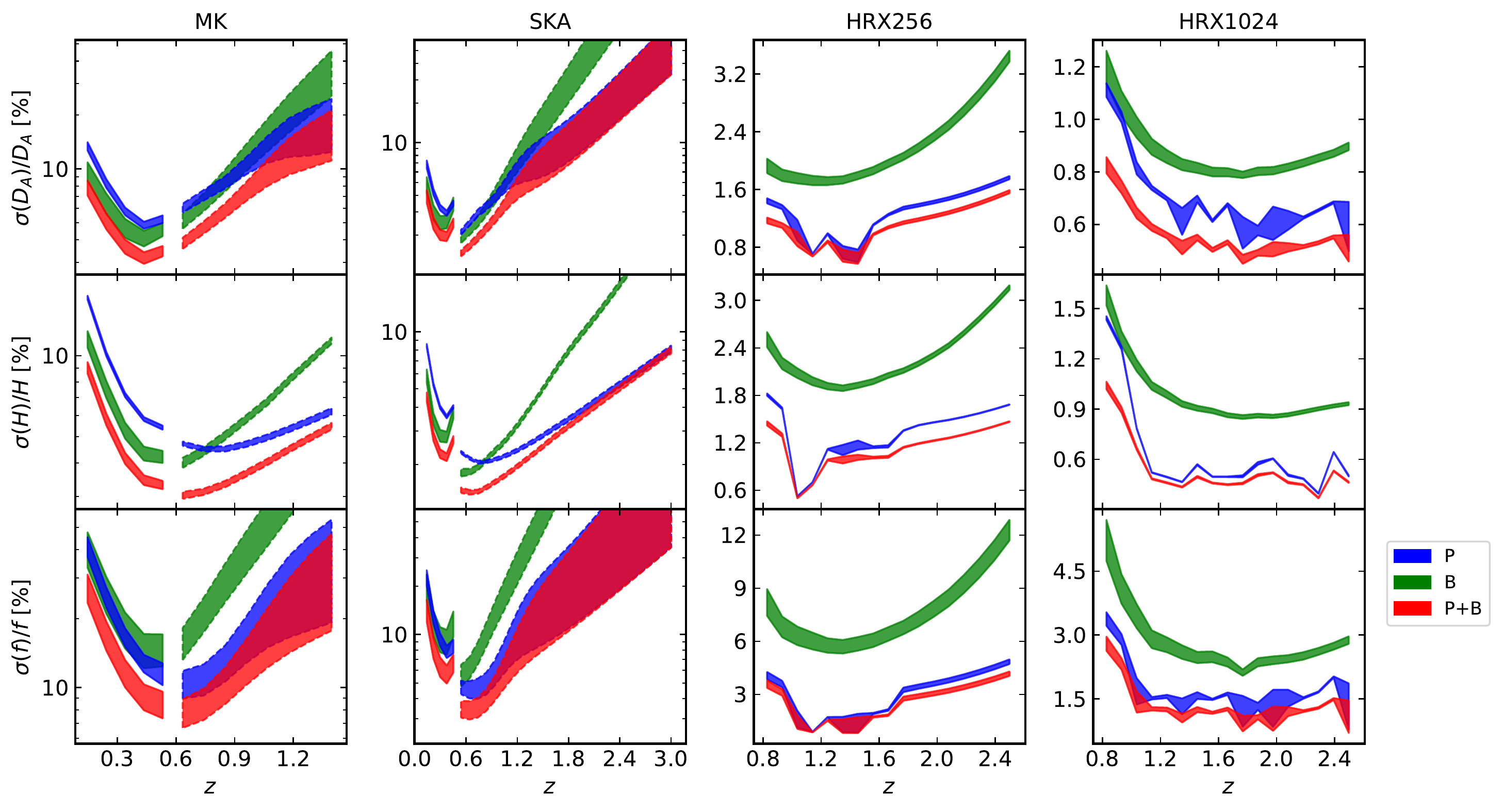}}
    \caption{Forecasts of marginalised relative errors on $f$ (bottom rows), $H$ (middle) and $D_A$ (top), from power spectrum (blue), bispectrum (green) and combined (red). For MeerKAT and SKAO (two left columns), the low-$z$ (solid lines) and high-$z$ (dashed) bands are shown. Upper ($\kparmin=0.01\;\Mpc$) and lower ($\kparmin=0$) boundaries of shaded regions correspond to the standard and idealised  foreground assumptions. Gaps in SD survey curves arise since information is taken at the midpoints of the last low-$z$ and first high-$z$ bins (taking into account the removal of the lowest Band 1 bins to avoid double counting).} 
\label{fig:distance_meas}
\end{figure*}

 \Cref{fig:distance_meas} displays the marginalised relative errors on the Hubble parameter, angular diameter distance and growth rate for all surveys considered here. \Planck priors are used for the $\Lambda$CDM parameters throughout this section. The shaded regions give the upper and lower bounds corresponding to  $\kparmin=0.01\;\Mpc$ and $\kparmin=0$ foreground cuts, discussed in \Cref{sec:surveys}.  The model assumed is the standard $\Lambda$CDM, where the mapping from $H(z)$ and $D_A(z)$ to the cosmological parameters, via the Jacobian transformation in \Cref{eq:Jacobian}, is not performed. In other words, the results presented in this section utilise the Fisher matrix of the initial parameter vector [\Cref{eq:params}], which is reduced to a $5+6N_z$ square matrix after marginalising over the nuisance parameters.

For the  SD surveys, the precision on  the BAO distance scale parameters is well below $10\%$ for most of the redshift range. In particular for MeerKAT, the constraints from the two correlators are mostly on the same level; for low redshifts, the bispectrum provides smaller errors (by a few percent). The reverse is observed for higher redshifts and specifically for most of the UHF band, where constraints from the power spectrum dominate. The inclusion of the bispectrum improves significantly the power spectrum results, especially for low redshifts, keeping the precision of the joint constraints at percent level for $0.4\lesssim z\lesssim 0.6$. SKAO forecasts follow a similar pattern, providing the tightest constraints around $z\sim0.6$, with errors larger at $z\gtrsim 1.5$. This is more evident for the angular diameter distance, where the high redshift bins of Band 1 offer no constraining power, for both correlators. 

The SD forecasts highlight the importance of including the bispectrum when measuring the BAO distance parameters, especially
for the low-$z$ bands, where instrumental noise allows  access to enough scales for the bispectrum signal to break parameter degeneracies, improving on the power spectrum constraints. This behaviour is reversed for $z\gtrsim0.8$, where the bispectrum signal does not make an important contribution. At the higher redshifts, the limitations of SD surveys (effects of beam and instrumental noise) degrade both statistics, especially the bispectrum. Angular diameter distance is afflicted the most, leading to very poor constraints for $z\gtrsim2$. On the other hand, the relative error on the Hubble parameter are mostly $\lesssim 10\%$,  reaching percent precision around $z\sim0.6$ for both SD surveys. 
The effect of the standard and idealised foreground cuts is minimal for the Hubble parameter, for the whole redshift range and both surveys. This is true as well for the angular diameter distance at low redshifts, while beyond $z\sim 1$ the absence of a foreground cut has a notable effect.  

Similarly to the cosmological parameters presented in the previous sections, HIRAX provides the most stringent constraints on $H$ and $D_A$, at sub-percent level for most of the redshift range, in the 256 and 1024 cases. Although IF surveys have the additional effect of the foreground wedge, they do not lose signal due to a wide beam at high $z$ and are able to access more of the smaller scales where signal is higher. The power spectrum consistently gives the tightest constraints, especially at high redshifts. Adding the bispectrum improves the results only marginally. For HIRAX-1024, the joint constraints  saturate gradually for $z\gtrsim 1.2$.  These forecasts are significantly better than for the SD surveys, in particular the high redshifts, where HIRAX  achieves almost a two orders of magnitude improvement over SKAO. 
The HIRAX errors on the BAO distance  parameters are basically unaffected by the presence of a foreground cut. This is perhaps not surprising, since IF mode surveys gain most of the signal from the intermediate and small scales. 

Note that the results consider the information from CMB, which is added as priors on the $\Lambda$CDM parameters, improving the signal and breaking various degeneracies between parameters. This leads to a significant improvement of the overall forecasts on $D_A$ and $H$, for all surveys considered, especially for the bispectrum results, which is almost an order of magnitude. The effect on the power spectrum is more limited (by a factor of 2-3), but still noticeable. The results without the priors are not shown here for brevity.

Constraints on the growth rate $f$ in the case of MeerKAT are modest: below $10\%$ for $0.4\lesssim z\lesssim0.8$. SKAO  reaches precision of $<4\%$ for $0.6\lesssim z\lesssim0.8$. For both SD surveys at higher redshift, the constraints worsen; for $z\gtrsim 1$ there is effectively no constraining power. At low $z$ the two correlators provide similar forecasts, and the bispectrum provides an improvement of $\sim10\%$, mainly due to the partial breaking of the well known degeneracies between $f$, $A_{\rm s}$ and linear bias. The tree-level power spectrum is insufficient to achieve this alone and therefore the contribution from the bispectrum can help, as  shown in \cite{Agarwal:2020lov,Ivanov:2021kcd}, at least for surveys where the three-point statistics has enough signal. At higher $z$, due to the SD beam, the contribution of the bispectrum is negligible and the forecasts are driven solely by the power spectrum. 

This is also the case for HIRAX:  the bispectrum contribution is minimal, but good enough to be competitive by itself, reaching a few percent precision for the entire redshift range. The power spectrum and the joint results provide a precision  $\lesssim3\%$ at all $z$ for HIRAX-256, while HIRAX-1024 doubles the precision at $z\gtrsim1$. 
The presence of a foreground cut has a small effect at all $z$ for HIRAX, with a larger effect for the SD surveys, especially at high $z$.

\subsection{Bias parameters}\label{sec:res_bias}

 \begin{figure*}[t]
\centering
\resizebox{\textwidth}{!}{\includegraphics{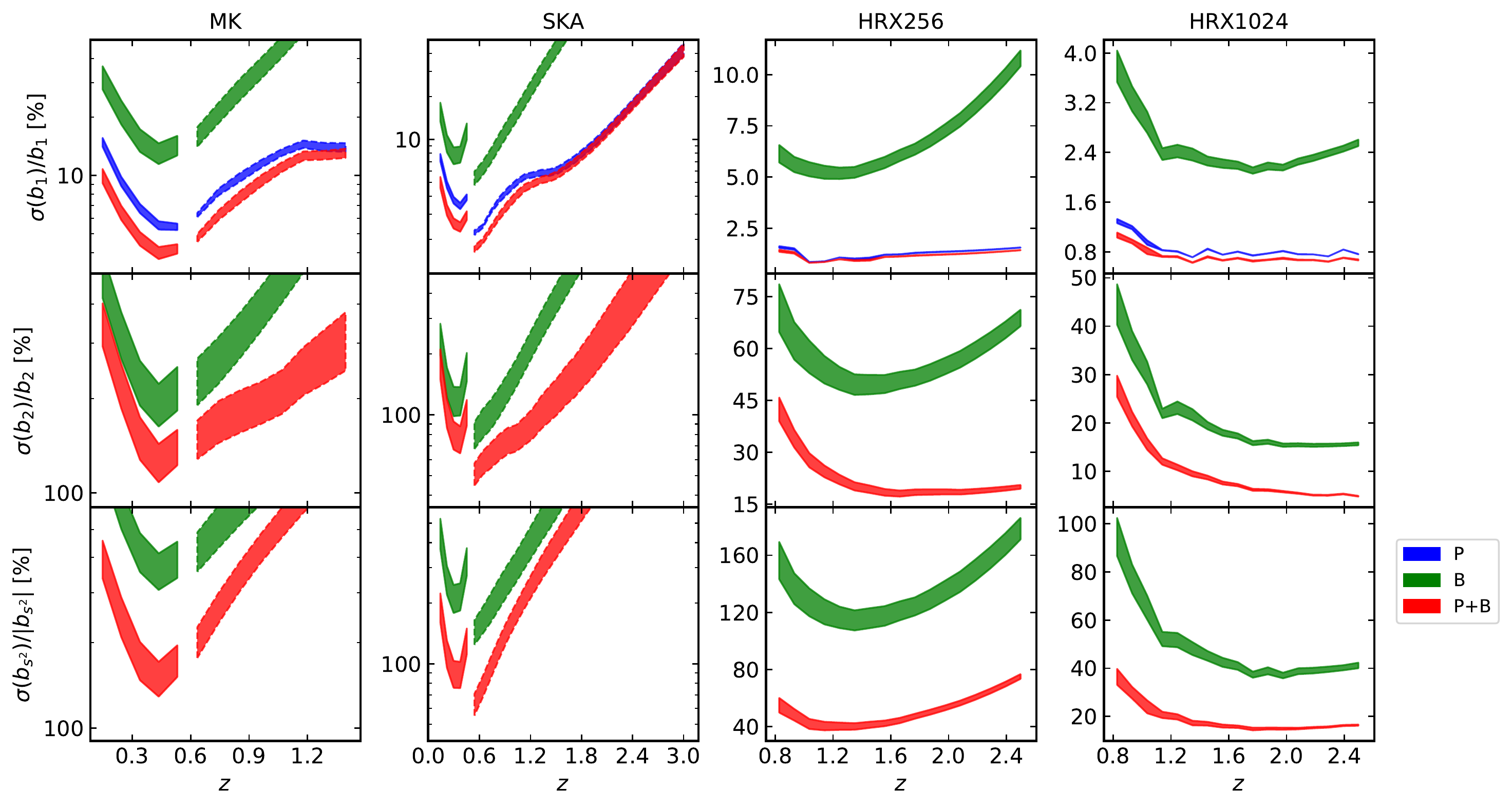}}
  \caption{As in \Cref{fig:distance_meas} but for the bias parameters described in \Cref{sec:halo_bias}.}
\label{fig:bias_meas}
\end{figure*}

Marginalised relative errors on the linear and second-order HI bias parameters (see \Cref{sec:halo_bias}) are presented in \Cref{fig:bias_meas}. 

The SD surveys have  modest constraining power on $b_1$, with maximum precision of a few percent achieved around $z\sim 0.5-0.6$. For both surveys, the constraints are driven by the power spectrum, with only a small contribution from the bispectrum. The joint errors are $<10\%$, until $z\sim1$ for both MeerKAT and SKAO. 
In the case of the second-order bias parameters, MeerKAT has no constraining power, while SKAO produces errors below $100\%$ only at low redshifts, based effectively only on the bispectrum.  SKAO is unusable for measuring $b_2$ and $b_{s^2}$. 

The second-order bias parameters appear only in the tree-level modelling of the bispectrum, so that for the scales and models considered here, the bispectrum is the sole contributor of signal. Indeed, the main effect expected from a bispectrum analysis is the measurement of bias parameters, as shown in \cite{Yankelevich2018,Chudaykin:2019ock,Agarwal:2020lov,Ivanov:2021kcd}. 
However, the SD bispectrum does not contain enough information to constrain $b_2$ and $b_{s^2}$, due to unresolved degeneracies between $b_1$ and $b_2$ and between $b_1$ and $b_{s^2}$. Nonetheless, the forecasts from the joint estimation yield a significant improvement (although still non-competitive), due to the precise SD constraints on $b_1$ by the power spectrum, which is enough to partially break degeneracies.

For HIRAX, the forecasts again display a great improvement over the SD surveys. The bispectrum plays even less of a role in  constraining linear bias than for the SD surveys. The precision on $b_1$ from the power spectrum and joint statistics is roughly constant over the redshift range at $\sim1\%$ for HIRAX-256 and sub-percent for HIRAX-1024. The bispectrum, despite its negligible effect in these constraints, has enough information by itself to constrain  $b_1$ at a few percent level,  reaching $\sim2.5\%$ at $z\gtrsim 1$ for HIRAX-1024. At high-$z$ we see that the bispectrum benefits from the larger $k_{\rm max}$ values, keeping the errors nearly constant. By contrast, the HIRAX-256 bispectrum shows rapidly growing errors at high $z$, the reason being that it is not a close-packed array like HIRAX-1024.
 
 For $b_2$, the HIRAX-256 bispectrum provides poor constraints, while for $b_{s^2}$ there is effectively no precision from the bispectrum. 
Fitting simultaneously the power spectrum and  bispectrum  improves the results, due to the high precision on $b_1$ contributed by the power spectrum, which breaks degeneracies between the bias parameters. Specifically, the  error on $b_2$ reaches $\sim20\%$ for $z\gtrsim 1.5$, while the precision on $b_{s^2}$ is within $40-80\%$, which is an improvement of a factor of $\sim2.5$. HIRAX-1024, with its close-packed array, significantly enhances precision, with a steady improvement towards high $z$, up to saturation. Bispectrum on its own reaches $\sim15\%$ on $b_{2}$ and $\sim40\%$ on $b_{s^2}$. The errors grow at lower $z$, to the point where HIRAX-1024 is unable to constrain  $b_{s^2}$. Adding the power spectrum information again  significantly improves  precision:  $\sim3\%$ at  high $z$ for $b_2$, and $\sim 20\%$, except at the lowest $z$ for  $b_{s^2}$. Once again, HIRAX-1024 provides significant better constraints than SKAO. 


The presence of a foreground cut has negligible effects on  bias parameter precision for the power spectrum and joint statistics, while there is  a more evident effect on the bispectrum, in particular on the $b_2$ constraints coming from the SD surveys and HIRAX-256.

Including the CMB measurements on the $\Lambda$CDM parameters improves the constraints on the linear bias significantly, mainly due to the breaking of degeneracies between $b_1$ and $A_{\rm s}$. The \Planck priors affect the power spectrum forecasts the most. MeerKAT low- and high-$z$ bins  show a $\sim 40\%$ improvement, with $\sim 50\%$ for the intermediate bins. For SKAO, priors mainly affect the constraints for $z<2$, with an improvement of $\sim 10-20\%$. The effect of CMB priors on the bispectrum contraints is less but still noticeable: $\sim 40-60\%$ for MeerKAT, mainly at $z\lesssim 1$; $\sim 15-20\%$ for SKAO at $z<1.5$. The remaining redshift bins are unaffected by CMB priors, mostly due to the low bispectrum signal at high-$z$ in SD surveys. The improvements from the joint power spectrum and bispectrum signal on $b_1$ are affected the least: $\sim 10-15\%$, with a maximum around $z\sim0.5-0.6$, for MeerKAT; a few percent at $z<1.5$ for SKAO. 
The improvement is minimal for both IF arrays, and both correlators.  For HIRAX-256 on $b_1$, it  is $1-5\%$ for the power spectrum and joint signal, while the bispectrum results improve by $3-10\%$, the best at low $z$. For HIRAX-1024, there is a marginal few-percent improvement for all $z$ and both correlators. 

On the other hand, $b_{2}$ and $b_{s^2}$ show minimal improvement for all surveys considered here and both correlators, since $b_2$ and $b_{s^2}$ are least correlated with cosmological parameters. Most of the signal is from the bispectrum, while the inclusion of priors mainly benefit the power spectrum via breaking parameter degeneracies.

\section{Conclusions}\label{sec:conc}

In this work we examine the potential of upcoming HI intensity mapping surveys,  using single-dish and interferometer modes, in constraining cosmology. These surveys enable us to probe the high-redshift Universe across wide sky area, increasing by a significant amount the observed volumes and therefore the cosmological signal of the chosen summary statistics. This could improve current measurements on cosmological and nuisance parameters. The question we try to address is: how much information is contained in the redshift space two- and three-point statistics of HI IM experiments, to constrain cosmological quantities?

The complete analytical tree-level model is used for the power spectrum and bispectrum, valid up to linear/quasi-linear clustering scales. For the power spectrum of the underlying matter field we use the non-linear treatment of the HMCode halo model \cite{Mead2020}, while for the matter bispectrum we use the tree-level prediction from standard perturbation theory. The general clustering bias prescription (\Cref{sec:halo_bias}) and  redshift-space mapping (\Cref{sec:RSDmodel}) are used up to the lowest non-vanishing order, \ie up to leading (power spectrum) and second order (bispectrum). Additionally, the FoG effect is included via an exponential damping factor [\Cref{eq:fog_PS,,eq:fog_BS}], and the AP effect  is incorporated in the standard way (\Cref{sec:APeffect}). The modelling used here is a consistent approach for both statistics, since we keep the analysis well within the perturbative regime by only considering scales that satisfy $k\leq k_{\rm max}$ [\Cref{kmax}]. At low redshift, where  non-linearity is stronger (\ie smaller $ k_{\rm max}$), we venture marginally into the quasi-linear regime. Due to this we also consider theoretical uncertainties in our analysis (\Cref{sec:theoretical_er}). The addition of the theoretical errors to the covariance matrix takes into account the effect of neglecting higher-order effects from the chosen model, which in turn makes the parameter forecasts insensitive to the choice of the $k_{\rm max}$ value. We also take into account instrumental effects: telescope beam,  instrumental thermal noise,  foreground avoidance via radial and wedge cuts (see \Cref{sec:surveys}).

The analysis uses the Fisher matrix formalism to forecast constraints on cosmological parameters, modified gravity, dark energy, distance measurements and clustering bias coefficients, marginalising over FoG and stochastic bias parameters. We employ the signals from the HI power spectrum, HI bispectrum, and their combination. The results from the HI surveys are also combined with \Planck constraints on the $\Lambda$CDM parameters, in the form of priors. The main results of this work are:

\begin{enumerate}[(1)]

    \item Upcoming HI IM surveys with a packed-array interferometer like HIRAX-1024, in combination with \Planck measurements, could improve significantly the precision on $\Lambda$CDM parameters, reaching sub-percent levels (see \Cref{table:results_IF}). 
    
    \item For the {SD} surveys, the power spectrum and bispectrum in redshift space have similar constraining power on cosmological parameters, with the power spectrum being marginally better (\Cref{fig:LCDM_triangle_plots}). Consequently, combining the information of the two correlators has a minimal impact on cosmological constraints relative to considering the power spectrum alone. For HIRAX surveys,  the bispectrum contribution can be neglected (\Cref{table:results_IF}). 
    
    \item  Synergy between HI surveys and CMB data is crucial for the SD surveys, in order to achieve stringent cosmological constraints. Only the summed signal from both SKAO bands can provide a meaningful improvement over \Planck measurements, while MeerKAT offers negligible  constraining power (\Cref{fig:LCDM_triangle_plots}).
    
    \item  For the SD surveys the bispectrum contribution is larger, but still below the power spectrum. However, once the priors are considered, the benefit from combining the signal of the two correlators is modest (\Cref{table:results_SD}).
    
    \item Combining HI surveys with CMB observations delivers strong potential for constraining the growth index $\gamma$. HIRAX-1024 could provide percent precision measurements from the combined signal of both correlators, while SKAO achieves $\sim5\%$ (\Cref{table:results_IF}). The bispectrum contribution is more important for SD surveys in constraining $\gamma$, while for IF experiments it is negligible (\Cref{fig:gCDM_triangle_plots}).
    
    \item The dark energy equation-of-state parameters $w_0$ and $w_a$ are not well constrained by the SD mode surveys. By contrast, HIRAX-1024 could constrain them with few-percent precision (\Cref{fig:wCDM_triangle_plots}).
    
    \item The power spectrum and bispectrum of the SD surveys provide similar forecasts, and considering both improves the dark energy FoM by a factor $\sim 1.5-2$ relative to the power spectrum alone. The FoM of HIRAX is mainly driven by the power spectrum signal (see bottom rows of \Cref{table:results_SD,,table:results_IF}).
    
    \item The combination of HIRAX and \Planck reaches sub-percent precision on $D_A(z)$ and $H(z)$  for the combined signal  (\Cref{fig:distance_meas}). This could be important for elucidating the nature of the so-called `Hubble tension' \cite{DiValentino:2021izs}. 
    
    \item For the SD surveys,  constraints on $D_A(z)$ and $H(z)$ are at $3-10\%$ precision for MeerKAT (except for $D_A$ at $z\gtrsim 1$), while SKAO achieves $1-2\%$ over the low  $z$ of Band 1.
    
    \item In the case of the growth rate $f(z)$, HIRAX can produce few-percent precision for the entire redshift range. SKAO can only do this in the low redshift slices (\Cref{fig:distance_meas}).
    
    \item An accurate assessment of the potential of future HI IM surveys to constrain cosmology, requires  precision 
    on the linear clustering bias parameter.  HIRAX can achieve $<2\%$ precision on $b_1$ using the HI power spectrum;  the bispectrum alone reaches few-percent precision, rendering its contribution negligible. A similar trend is shown for the SD surveys, where few-percent precision is achieved only at low redshifts.
    
    \item The quadratic bias parameters, within the tree-level analysis used here, can be constrained only by the HI bispectrum. Thus, the SD surveys, due to their intrinsic limitations (see \Cref{sec:surveys}), are unable to provide useful constraints. On the other hand, HIRAX-1024, which as an IF survey achieves stronger bispectrum constraints, can deliver precision of $\sim5\%$ on $b_2$ and $\sim 18\%$,  on $b_{s^2}$  (\Cref{fig:bias_meas}).
    
    \item The standard and idealised values chosen for the radial mode cut-off in the foreground avoidance, have a minimal effect on most parameter forecasts and surveys considered here. In particular, the presence of a foreground cut seems to have a moderate effect mainly on the constraints of the growth rate and the non-linear bias, coming from the large redshift bins of the SD surveys. For these surveys and redshifts, this is also true for the angular diameter distance. We additionally checked the result of a harder foreground cut $\kparmin=0.05\;\Mpc$, finding a negligible change in the errors, compared to the idealised case, for most of the parameters and surveys. More precisely, the only notable change is a $5-10\%$ increase in the angular diameter distance and growth rate errors from the high redshift bins of the SD surveys, while for the non-linear bias parameter the increase is within $2-5\%$.

\end{enumerate}

The superior performance that is forecast for HIRAX is not unexpected. Constraining power on cosmological parameters and on the BAO distance and growth rate functions, relies on access to the higher signal on smaller scales. An interferometer such as HIRAX covers these scales particularly well, whereas SD surveys progressively lose these scales as redshift increases, due to the telescope beam \cite{Bull2015,Karagiannis:2020dpq}. Indeed, HIRAX is designed as a BAO intensity mapping `machine' \cite{Crichton:2021hlc}. By contrast, the SKAO interferometer was not designed with HI intensity mapping in mind, so that SKAO is better in single-dish mode for intensity mapping cosmology \cite{Bacon:2018dui}. For cosmological constraints that require access to very large scales, such as measuring the turnover of the power spectrum \cite{Cunnington:2022ryj}, or probing local primordial non-Gaussianity via scale-dependent bias \cite{Viljoen:2021ypp}, SKAO in SD mode is more capable than an interferometer like HIRAX \cite{Karagiannis:2020dpq}.
Finally, we should point out that pilot intensity mapping surveys (with the associated data pipeline construction) are already underway on the SKAO precursor MeerKAT \cite{Li:2020bcr,Paul:2020ank,Matshawule:2020fjz,Wang:2020lkn,Cunnington:2022uzo}, while the HIRAX 256-dish precursor is not yet constructed.

\[\]

\section*{Acknowledgements}

The authors are supported by the South African Radio Astronomy Observatory
and the National Research Foundation (Grant No. 75415).

\clearpage
\appendix
\section{Baseline distribution for HIRAX} \label{app:baseline}

An {idealised theoretical model} is proposed in \cite{PUMA2018}  for the  baseline density of close-packed square (HIRAX-like) and hexagonal (PUMA-like) arrays.
First,  the image-plane density is related to the physical density:
\begin{equation}
\label{eq:nu_z}
n_{\rm b}({u},z)= \lambda(z)^2\, n_{\rm b}^{\rm phys}(L)\quad \mbox{where}\quad L= u\,\lambda\,. 
 \end{equation}
Then a fitting formula for the model of the physical baseline density is given as  \cite{PUMA2018}
 \begin{equation}
 \label{eq:nu_phys}
n_{\rm b}^{\rm phys}(L) = \left(\frac{N_{\rm s}}{D_{\rm dish}}\right)^2\,\frac{a_1 +a_2\,(L/L_{\rm s})}{1+a_3\,(L/L_{\rm s})^{a_4}}\,\exp\left[-\left({L\over L_{\rm s}}\right)^{a_5}\right],
\end{equation}
where $L_{\rm s} = N_{\rm s}\,D_{\rm dish}$ and $N_{\rm s}^2 =  N_{\rm dish}$. For HIRAX, $N_{\rm s}=16$ and 32 in the early and full stages respectively. 
The maximum baseline is the diagonal of the square array: $D_{\rm max}\approx {\sqrt{2}\, L_{\rm s}}$ \cite{Jolicoeur:2020eup}.
The parameters in \Cref{eq:nu_phys} for a square closely-packed array like HIRAX are \cite{PUMA2018}
\begin{equation}
a_I=\big(0.4847\,,\, -0.3300\,,\, 1.3157 \,,\, 1.5974 \,,\, 6.8390 \big).
\end{equation}

\begin{figure}[!h]
    \centering
    \includegraphics[width=0.48\linewidth]{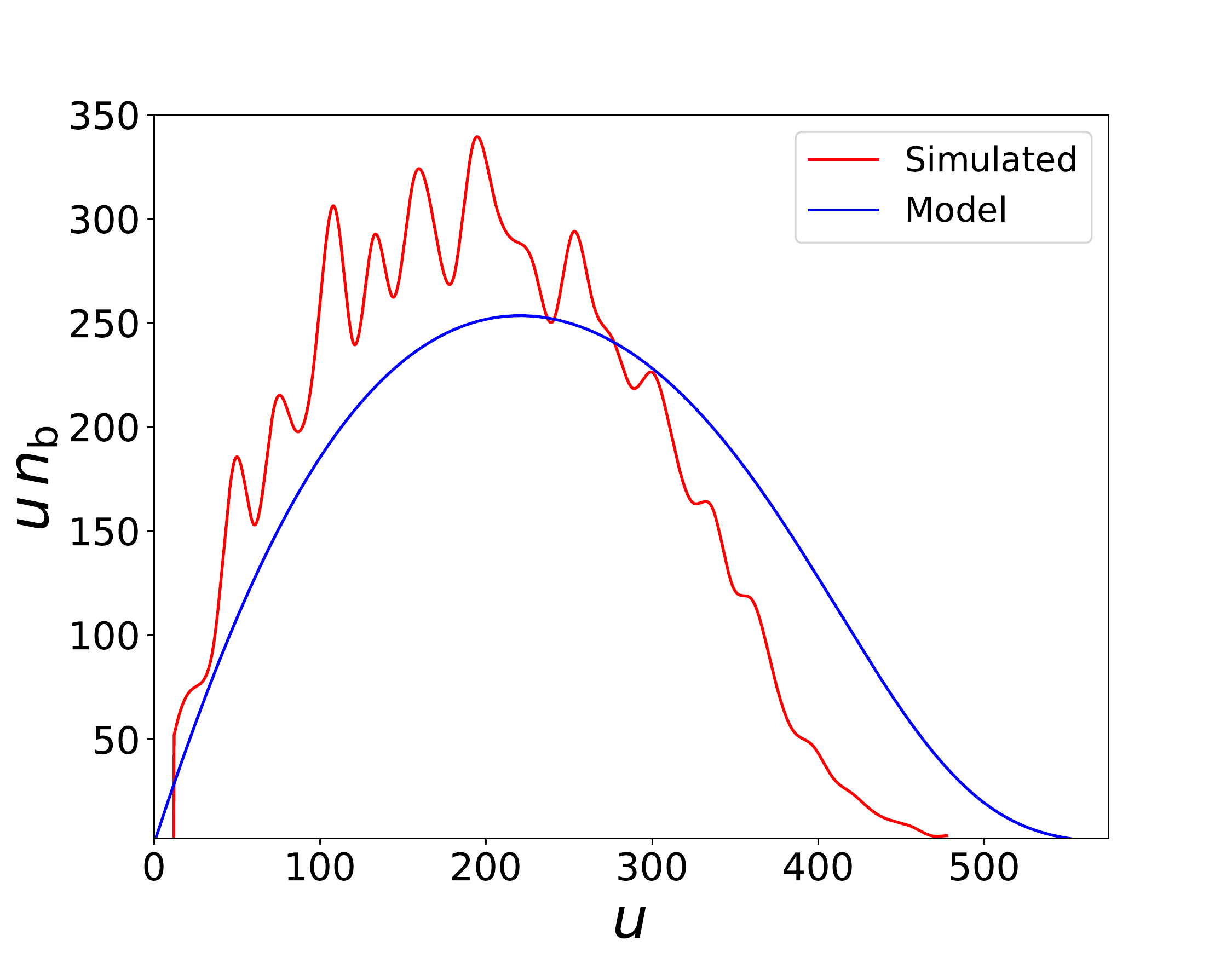}
    \quad
    \includegraphics[width=0.48\linewidth]{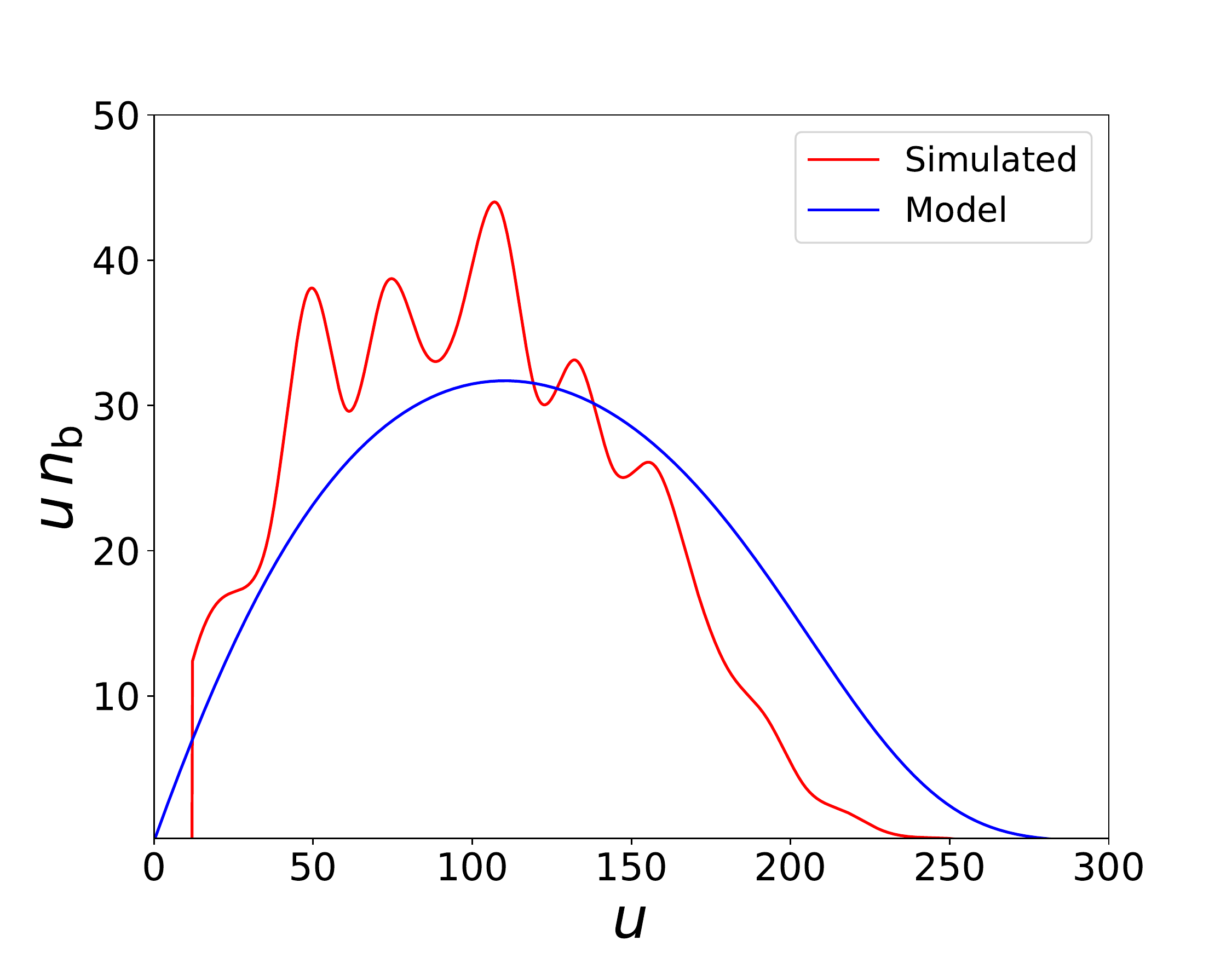}
    \caption{
     Comparison between the simulated  baseline density (red) and the idealised model \Cref{eq:nu_phys} (blue), for HIRAX-1024 ({\em left}) and HIRAX-256 ({\em right}). }
    \label{fig:baseline_density}
\end{figure}

Instead of using an idealised model, we use the results from simulations of the HIRAX array \cite{Crichton:2021hlc}.
These simulations 
are shown by the red curves in \Cref{fig:baseline_density}  for 1024 (left) and 256 (right) dishes.
The baseline density that follows from the fitting formula \Cref{eq:nu_phys} is the blue curve. It is apparent that the fitting formula (blue) does not provide a very good match to the simulations (red).
{Different models of the baseline density $n_{\rm b}$ should each satisfy the constraint that the total number of baselines is $N_{\rm dish}(N_{\rm dish}-1)/2\approx N_{\rm dish}^2/2$. This implies that
\begin{equation}\label{bla}
 \int {\rm d}u\,u\, n_{\rm b}\approx {N_{\rm dish}^2\over 4\pi}\,,  
\end{equation}
where we used ${\rm d}^2\bm{u}=2\pi\,u\,{\rm d}u$, assuming azimuthal symmetry. The relation \Cref{bla} is satisfied by the simulated (red) and idealised (blue) curves in \Cref{fig:baseline_density}.}

\clearpage
\bibliography{references}

\providecommand{\href}[2]{#2}\begingroup\raggedright\begin{thebibliography}{100}

\bibitem{Planck2018_cosmo}
{Planck Collaboration}, N.~{Aghanim}, Y.~{Akrami}, M.~{Ashdown}, J.~{Aumont},
  C.~{Baccigalupi} et~al., \emph{{Planck 2018 results. VI. Cosmological
  parameters}}, \href{https://doi.org/10.1051/0004-6361/201833910}{\emph{\aap}
  {\bfseries 641} (2020) A6}
  [\href{https://arxiv.org/abs/1807.06209}{{\ttfamily 1807.06209}}].

\bibitem{DESI2016}
{DESI Collaboration}, A.~{Aghamousa}, J.~{Aguilar}, S.~{Ahlen}, S.~{Alam},
  L.~E. {Allen} et~al., \emph{{The DESI Experiment Part I: Science,Targeting,
  and Survey Design}}, {\emph{ArXiv e-prints} (2016) }
  [\href{https://arxiv.org/abs/1611.00036}{{\ttfamily 1611.00036}}].

\bibitem{Euclid:2019clj}
{\scshape Euclid} collaboration, A.~Blanchard et~al., \emph{{Euclid
  preparation: VII. Forecast validation for Euclid cosmological probes}},
  \href{https://doi.org/10.1051/0004-6361/202038071}{\emph{Astron. Astrophys.}
  {\bfseries 642} (2020) A191}
  [\href{https://arxiv.org/abs/1910.09273}{{\ttfamily 1910.09273}}].

\bibitem{LSSTDarkEnergyScience:2018jkl}
{\scshape LSST Dark Energy Science} collaboration, R.~Mandelbaum et~al.,
  \emph{{The LSST Dark Energy Science Collaboration (DESC) Science Requirements
  Document}},  \href{https://arxiv.org/abs/1809.01669}{{\ttfamily 1809.01669}}.

\bibitem{Bacon:2018dui}
{\scshape SKA} collaboration, D.~J. Bacon et~al., \emph{{Cosmology with Phase 1
  of the Square Kilometre Array: Red Book 2018: Technical specifications and
  performance forecasts}},
  \href{https://doi.org/10.1017/pasa.2019.51}{\emph{Publ. Astron. Soc.
  Austral.} {\bfseries 37} (2020) e007}
  [\href{https://arxiv.org/abs/1811.02743}{{\ttfamily 1811.02743}}].

\bibitem{Crichton:2021hlc}
D.~Crichton et~al., \emph{{The Hydrogen Intensity and Real-time Analysis
  eXperiment: 256-Element Array Status and Overview}},
  \href{https://doi.org/10.1117/1.JATIS.8.1.011019}{\emph{J. Astron. Telesc.
  Instrum. Syst.} {\bfseries 8} (2022) 011019}
  [\href{https://arxiv.org/abs/2109.13755}{{\ttfamily 2109.13755}}].

\bibitem{Santos:2015gra}
M.~G. Santos et~al., \emph{{Cosmology from a SKA HI intensity mapping survey}},
  \href{https://doi.org/10.22323/1.215.0019}{\emph{PoS} {\bfseries AASKA14}
  (2015) 019} [\href{https://arxiv.org/abs/1501.03989}{{\ttfamily
  1501.03989}}].

\bibitem{Ivanov:2021kcd}
M.~M. Ivanov, O.~H.~E. Philcox, T.~Nishimichi, M.~Simonovi\'c, M.~Takada and
  M.~Zaldarriaga, \emph{{Precision analysis of the redshift-space galaxy
  bispectrum}}, \href{https://doi.org/10.1103/PhysRevD.105.063512}{\emph{Phys.
  Rev. D} {\bfseries 105} (2022) 063512}
  [\href{https://arxiv.org/abs/2110.10161}{{\ttfamily 2110.10161}}].

\bibitem{Philcox:2021kcw}
O.~H.~E. Philcox and M.~M. Ivanov, \emph{{BOSS DR12 full-shape cosmology:
  \ensuremath{\Lambda}CDM constraints from the large-scale galaxy power
  spectrum and bispectrum monopole}},
  \href{https://doi.org/10.1103/PhysRevD.105.043517}{\emph{Phys. Rev. D}
  {\bfseries 105} (2022) 043517}
  [\href{https://arxiv.org/abs/2112.04515}{{\ttfamily 2112.04515}}].

\bibitem{DAmico:2022gki}
G.~D'Amico, M.~Lewandowski, L.~Senatore and P.~Zhang, \emph{{Limits on
  primordial non-Gaussianities from BOSS galaxy-clustering data}},
  \href{https://arxiv.org/abs/2201.11518}{{\ttfamily 2201.11518}}.

\bibitem{Cabass:2022ymb}
G.~Cabass, M.~M. Ivanov, O.~H.~E. Philcox, M.~Simonovi\'c and M.~Zaldarriaga,
  \emph{{Constraints on Multi-Field Inflation from the BOSS Galaxy Survey}},
  \href{https://arxiv.org/abs/2204.01781}{{\ttfamily 2204.01781}}.

\bibitem{Veropalumbo:2022zfs}
A.~Veropalumbo, A.~Binetti, E.~Branchini, M.~Moresco, P.~Monaco, A.~Oddo
  et~al., \emph{{The galaxy 3-point correlation function: a methodological
  analysis}},  \href{https://arxiv.org/abs/2206.00672}{{\ttfamily 2206.00672}}.

\bibitem{Rizzo:2022lmh}
F.~Rizzo, C.~Moretti, K.~Pardede, A.~Eggemeier, A.~Oddo, E.~Sefusatti et~al.,
  \emph{{The Halo Bispectrum Multipoles in Redshift Space}},
  \href{https://arxiv.org/abs/2204.13628}{{\ttfamily 2204.13628}}.

\bibitem{Coulton:2022qbc}
W.~R. Coulton, F.~Villaescusa-Navarro, D.~Jamieson, M.~Baldi, G.~Jung,
  D.~Karagiannis et~al., \emph{{Quijote-PNG: Simulations of primordial
  non-Gaussianity and the information content of the matter field power
  spectrum and bispectrum}},
  \href{https://arxiv.org/abs/2206.01619}{{\ttfamily 2206.01619}}.

\bibitem{Jung:2022rtn}
G.~Jung, D.~Karagiannis, M.~Liguori, M.~Baldi, W.~R. Coulton, D.~Jamieson
  et~al., \emph{{Quijote-PNG: Quasi-maximum likelihood estimation of Primordial
  Non-Gaussianity in the non-linear dark matter density field}},
  \href{https://arxiv.org/abs/2206.01624}{{\ttfamily 2206.01624}}.

\bibitem{Philcox:2022frc}
O.~H.~E. Philcox, M.~M. Ivanov, G.~Cabass, M.~Simonovi\'c, M.~Zaldarriaga and
  T.~Nishimichi, \emph{{Cosmology with the Redshift-Space Galaxy Bispectrum
  Monopole at One-Loop Order}},
  \href{https://arxiv.org/abs/2206.02800}{{\ttfamily 2206.02800}}.

\bibitem{Karagiannis:2019jjx}
D.~Karagiannis, A.~Slosar and M.~Liguori, \emph{{Forecasts on Primordial
  non-Gaussianity from 21 cm Intensity Mapping experiments}},
  \href{https://doi.org/10.1088/1475-7516/2020/11/052}{\emph{JCAP} {\bfseries
  11} (2020) 052} [\href{https://arxiv.org/abs/1911.03964}{{\ttfamily
  1911.03964}}].

\bibitem{Karagiannis:2020dpq}
D.~Karagiannis, J.~Fonseca, R.~Maartens and S.~Camera, \emph{{Probing
  primordial non-Gaussianity with the power spectrum and bispectrum of future
  21 cm intensity maps}},
  \href{https://doi.org/10.1016/j.dark.2021.100821}{\emph{Phys. Dark Univ.}
  {\bfseries 32} (2021) 100821}
  [\href{https://arxiv.org/abs/2010.07034}{{\ttfamily 2010.07034}}].

\bibitem{Bull2015}
P.~{Bull}, P.~G. {Ferreira}, P.~{Patel} and M.~G. {Santos}, \emph{{Late-time
  Cosmology with 21\,cm Intensity Mapping Experiments}},
  \href{https://doi.org/10.1088/0004-637X/803/1/21}{\emph{\apj} {\bfseries 803}
  (2015) 21} [\href{https://arxiv.org/abs/1405.1452}{{\ttfamily 1405.1452}}].

\bibitem{Spinelli:2021emp}
M.~Spinelli, I.~P. Carucci, S.~Cunnington, S.~E. Harper, M.~O. Irfan,
  J.~Fonseca et~al., \emph{{SKAO HI intensity mapping: blind foreground
  subtraction challenge}},
  \href{https://doi.org/10.1093/mnras/stab3064}{\emph{Mon. Not. Roy. Astron.
  Soc.} {\bfseries 509} (2021) 2048}
  [\href{https://arxiv.org/abs/2107.10814}{{\ttfamily 2107.10814}}].

\bibitem{Bernardeau2002}
F.~{Bernardeau}, S.~{Colombi}, E.~{Gazta{\~n}aga} and R.~{Scoccimarro},
  \emph{{Large-scale structure of the Universe and cosmological perturbation
  theory}},
  \href{https://doi.org/10.1016/S0370-1573(02)00135-7}{\emph{\physrep}
  {\bfseries 367} (2002) 1}
  [\href{https://arxiv.org/abs/astro-ph/0112551}{{\ttfamily
  astro-ph/0112551}}].

\bibitem{Jackson1972}
J.~C. {Jackson}, \emph{{A critique of Rees's theory of primordial gravitational
  radiation}}, \href{https://doi.org/10.1093/mnras/156.1.1P}{\emph{\mnras}
  {\bfseries 156} (1972) 1P} [\href{https://arxiv.org/abs/0810.3908}{{\ttfamily
  0810.3908}}].

\bibitem{CAMB}
A.~{Lewis}, A.~{Challinor} and A.~{Lasenby}, \emph{{Efficient Computation of
  Cosmic Microwave Background Anisotropies in Closed Friedmann-Robertson-Walker
  Models}}, \href{https://doi.org/10.1086/309179}{\emph{\apj} {\bfseries 538}
  (2000) 473} [\href{https://arxiv.org/abs/astro-ph/9911177}{{\ttfamily
  astro-ph/9911177}}].

\bibitem{Assassi2014}
V.~{Assassi}, D.~{Baumann}, D.~{Green} and M.~{Zaldarriaga},
  \emph{{Renormalized halo bias}},
  \href{https://doi.org/10.1088/1475-7516/2014/08/056}{\emph{\jcap} {\bfseries
  8} (2014) 056} [\href{https://arxiv.org/abs/1402.5916}{{\ttfamily
  1402.5916}}].

\bibitem{Senatore2014}
L.~{Senatore}, \emph{{Bias in the effective field theory of large scale
  structures}},
  \href{https://doi.org/10.1088/1475-7516/2015/11/007}{\emph{\jcap} {\bfseries
  11} (2015) 007} [\href{https://arxiv.org/abs/1406.7843}{{\ttfamily
  1406.7843}}].

\bibitem{Mirbabayi2014}
M.~{Mirbabayi}, F.~{Schmidt} and M.~{Zaldarriaga}, \emph{{Biased tracers and
  time evolution}},
  \href{https://doi.org/10.1088/1475-7516/2015/07/030}{\emph{\jcap} {\bfseries
  7} (2015) 030} [\href{https://arxiv.org/abs/1412.5169}{{\ttfamily
  1412.5169}}].

\bibitem{Desjacques2016}
V.~{Desjacques}, D.~{Jeong} and F.~{Schmidt}, \emph{{Large-scale galaxy bias}},
  \href{https://doi.org/10.1016/j.physrep.2017.12.002}{\emph{\physrep}
  {\bfseries 733} (2018) 1} [\href{https://arxiv.org/abs/1611.09787}{{\ttfamily
  1611.09787}}].

\bibitem{Dekel1998}
A.~{Dekel} and O.~{Lahav}, \emph{{Stochastic Nonlinear Galaxy Biasing}},
  \href{https://doi.org/10.1086/307428}{\emph{\apj} {\bfseries 520} (1999) 24}
  [\href{https://arxiv.org/abs/astro-ph/9806193}{{\ttfamily
  astro-ph/9806193}}].

\bibitem{Taruya1998}
A.~{Taruya} and J.~{Soda}, \emph{{Stochastic Biasing and the Galaxy-Mass
  Density Relation in the Weakly Nonlinear Regime}},
  \href{https://doi.org/10.1086/307612}{\emph{\apj} {\bfseries 522} (1999) 46}
  [\href{https://arxiv.org/abs/astro-ph/9809204}{{\ttfamily
  astro-ph/9809204}}].

\bibitem{Matsubara1999}
T.~{Matsubara}, \emph{{Stochasticity of Bias and Nonlocality of Galaxy
  Formation: Linear Scales}}, \href{https://doi.org/10.1086/307931}{\emph{\apj}
  {\bfseries 525} (1999) 543}
  [\href{https://arxiv.org/abs/astro-ph/9906029}{{\ttfamily
  astro-ph/9906029}}].

\bibitem{Seljak2000}
U.~Seljak, \emph{{Analytic model for galaxy and dark matter clustering}},
  \href{https://doi.org/10.1046/j.1365-8711.2000.03715.x}{\emph{Mon. Not. Roy.
  Astron. Soc.} {\bfseries 318} (2000) 203}
  [\href{https://arxiv.org/abs/astro-ph/0001493}{{\ttfamily
  astro-ph/0001493}}].

\bibitem{Peacock2000}
J.~Peacock and R.~Smith, \emph{{Halo occupation numbers and galaxy bias}},
  \href{https://doi.org/10.1046/j.1365-8711.2000.03779.x}{\emph{Mon. Not. Roy.
  Astron. Soc.} {\bfseries 318} (2000) 1144}
  [\href{https://arxiv.org/abs/astro-ph/0005010}{{\ttfamily
  astro-ph/0005010}}].

\bibitem{Scoccimarro2000}
R.~Scoccimarro, R.~K. Sheth, L.~Hui and B.~Jain, \emph{{How many galaxies fit
  in a halo? Constraints on galaxy formation efficiency from spatial
  clustering}}, \href{https://doi.org/10.1086/318261}{\emph{Astrophys. J.}
  {\bfseries 546} (2001) 20}
  [\href{https://arxiv.org/abs/astro-ph/0006319}{{\ttfamily
  astro-ph/0006319}}].

\bibitem{Villaescusa2014}
F.~Villaescusa-Navarro, M.~Viel, K.~K. Datta and T.~R. Choudhury,
  \emph{{Modeling the neutral hydrogen distribution in the post-reionization
  Universe: intensity mapping}},
  \href{https://doi.org/10.1088/1475-7516/2014/09/050}{\emph{JCAP} {\bfseries
  09} (2014) 050} [\href{https://arxiv.org/abs/1405.6713}{{\ttfamily
  1405.6713}}].

\bibitem{Castorina2016}
E.~{Castorina} and F.~{Villaescusa-Navarro}, \emph{{On the spatial distribution
  of neutral hydrogen in the Universe: bias and shot-noise of the HI power
  spectrum}}, \href{https://doi.org/10.1093/mnras/stx1599}{\emph{\mnras}
  {\bfseries 471} (2017) 1788}
  [\href{https://arxiv.org/abs/1609.05157}{{\ttfamily 1609.05157}}].

\bibitem{Tinker2008}
J.~{Tinker}, A.~V. {Kravtsov}, A.~{Klypin}, K.~{Abazajian}, M.~{Warren},
  G.~{Yepes} et~al., \emph{{Toward a Halo Mass Function for Precision
  Cosmology: The Limits of Universality}},
  \href{https://doi.org/10.1086/591439}{\emph{\apj} {\bfseries 688} (2008) 709}
  [\href{https://arxiv.org/abs/0803.2706}{{\ttfamily 0803.2706}}].

\bibitem{Cooray2002}
A.~{Cooray} and R.~{Sheth}, \emph{{Halo models of large scale structure}},
  \href{https://doi.org/10.1016/S0370-1573(02)00276-4}{\emph{\physrep}
  {\bfseries 372} (2002) 1}
  [\href{https://arxiv.org/abs/astro-ph/0206508}{{\ttfamily
  astro-ph/0206508}}].

\bibitem{Villaescusa-Navarro:2015cca}
F.~Villaescusa-Navarro, P.~Bull and M.~Viel, \emph{{Weighing neutrinos with
  cosmic neutral hydrogen}},
  \href{https://doi.org/10.1088/0004-637X/814/2/146}{\emph{Astrophys. J.}
  {\bfseries 814} (2015) 146}
  [\href{https://arxiv.org/abs/1507.05102}{{\ttfamily 1507.05102}}].

\bibitem{Villaescusa-Navarro:2015zaa}
F.~Villaescusa-Navarro et~al., \emph{{Neutral hydrogen in galaxy clusters:
  impact of AGN feedback and implications for intensity mapping}},
  \href{https://doi.org/10.1093/mnras/stv2904}{\emph{Mon. Not. Roy. Astron.
  Soc.} {\bfseries 456} (2016) 3553}
  [\href{https://arxiv.org/abs/1510.04277}{{\ttfamily 1510.04277}}].

\bibitem{Pontzen:2008mx}
A.~Pontzen, F.~Governato, M.~Pettini, C.~Booth, G.~Stinson, J.~Wadsley et~al.,
  \emph{{Damped Lyman Alpha Systems in Galaxy Formation Simulations}},
  \href{https://doi.org/10.1111/j.1365-2966.2008.13782.x}{\emph{Mon. Not. Roy.
  Astron. Soc.} {\bfseries 390} (2008) 1349}
  [\href{https://arxiv.org/abs/0804.4474}{{\ttfamily 0804.4474}}].

\bibitem{Marin2010}
F.~A. {Mar{\'\i}n}, N.~Y. {Gnedin}, H.-J. {Seo} and A.~{Vallinotto},
  \emph{{Modeling the Large-scale Bias of Neutral Hydrogen}},
  \href{https://doi.org/10.1088/0004-637X/718/2/972}{\emph{\apj} {\bfseries
  718} (2010) 972} [\href{https://arxiv.org/abs/0911.0041}{{\ttfamily
  0911.0041}}].

\bibitem{Tinker2010}
J.~L. {Tinker}, B.~E. {Robertson}, A.~V. {Kravtsov}, A.~{Klypin}, M.~S.
  {Warren}, G.~{Yepes} et~al., \emph{{The Large-scale Bias of Dark Matter
  Halos: Numerical Calibration and Model Tests}},
  \href{https://doi.org/10.1088/0004-637X/724/2/878}{\emph{\apj} {\bfseries
  724} (2010) 878} [\href{https://arxiv.org/abs/1001.3162}{{\ttfamily
  1001.3162}}].

\bibitem{Lazeyras2016}
T.~{Lazeyras}, C.~{Wagner}, T.~{Baldauf} and F.~{Schmidt}, \emph{{Precision
  measurement of the local bias of dark matter halos}},
  \href{https://doi.org/10.1088/1475-7516/2016/02/018}{\emph{\jcap} {\bfseries
  2016} (2016) 018} [\href{https://arxiv.org/abs/1511.01096}{{\ttfamily
  1511.01096}}].

\bibitem{Baldauf2012}
T.~{Baldauf}, U.~{Seljak}, V.~{Desjacques} and P.~{McDonald}, \emph{{Evidence
  for quadratic tidal tensor bias from the halo bispectrum}},
  \href{https://doi.org/10.1103/PhysRevD.86.083540}{\emph{\prd} {\bfseries 86}
  (2012) 083540} [\href{https://arxiv.org/abs/1201.4827}{{\ttfamily
  1201.4827}}].

\bibitem{Sargent:1977}
W.~L.~W. {Sargent} and E.~L. {Turner}, \emph{{A statistical method for
  determining the cosmological density parameter from the redshifts of a
  complete sample of galaxies}},
  \href{https://doi.org/10.1086/182362}{\emph{Astrophys. J.} {\bfseries 212}
  (1977) L3}.

\bibitem{Kaiser1987}
N.~{Kaiser}, \emph{{Clustering in real space and in redshift space}},
  \href{https://doi.org/10.1093/mnras/227.1.1}{\emph{\mnras} {\bfseries 227}
  (1987) 1}.

\bibitem{Hamilton1998}
A.~J.~S. {Hamilton}, \emph{{Linear Redshift Distortions: a Review}},  in
  \emph{The Evolving Universe} (D.~{Hamilton}, ed.), vol.~231 of
  \emph{Astrophysics and Space Science Library}, p.~185, 1998,
  \href{https://arxiv.org/abs/astro-ph/9708102}{{\ttfamily astro-ph/9708102}},
  \href{https://doi.org/10.1007/978-94-011-4960-0_17}{DOI}.

\bibitem{Battye2013}
R.~A. {Battye}, I.~W.~A. {Browne}, C.~{Dickinson}, G.~{Heron}, B.~{Maffei} and
  A.~{Pourtsidou}, \emph{{H I intensity mapping: a single dish approach}},
  \href{https://doi.org/10.1093/mnras/stt1082}{\emph{\mnras} {\bfseries 434}
  (2013) 1239} [\href{https://arxiv.org/abs/1209.0343}{{\ttfamily 1209.0343}}].

\bibitem{McDonald2009}
P.~{McDonald} and A.~{Roy}, \emph{{Clustering of dark matter tracers:
  generalizing bias for the coming era of precision LSS}},
  \href{https://doi.org/10.1088/1475-7516/2009/08/020}{\emph{\jcap} {\bfseries
  8} (2009) 020} [\href{https://arxiv.org/abs/0902.0991}{{\ttfamily
  0902.0991}}].

\bibitem{Peacock1994}
J.~A. {Peacock} and S.~J. {Dodds}, \emph{{Reconstructing the Linear Power
  Spectrum of Cosmological Mass Fluctuations}},
  \href{https://doi.org/10.1093/mnras/267.4.1020}{\emph{\mnras} {\bfseries 267}
  (1994) 1020} [\href{https://arxiv.org/abs/astro-ph/9311057}{{\ttfamily
  astro-ph/9311057}}].

\bibitem{Ballinger1996}
W.~E. {Ballinger}, J.~A. {Peacock} and A.~F. {Heavens}, \emph{{Measuring the
  cosmological constant with redshift surveys}},
  \href{https://doi.org/10.1093/mnras/282.3.877}{\emph{\mnras} {\bfseries 282}
  (1996) 877} [\href{https://arxiv.org/abs/astro-ph/9605017}{{\ttfamily
  astro-ph/9605017}}].

\bibitem{Schmidt2015}
F.~{Schmidt}, \emph{{Towards a self-consistent halo model for the nonlinear
  large-scale structure}},
  \href{https://doi.org/10.1103/PhysRevD.93.063512}{\emph{\prd} {\bfseries 93}
  (2016) 063512} [\href{https://arxiv.org/abs/1511.02231}{{\ttfamily
  1511.02231}}].

\bibitem{Scoccimarro1999}
R.~{Scoccimarro}, H.~M.~P. {Couchman} and J.~A. {Frieman}, \emph{{The
  Bispectrum as a Signature of Gravitational Instability in Redshift Space}},
  \href{https://doi.org/10.1086/307220}{\emph{\apj} {\bfseries 517} (1999) 531}
  [\href{https://arxiv.org/abs/astro-ph/9808305}{{\ttfamily
  astro-ph/9808305}}].

\bibitem{Gagrani:2016rfy}
P.~Gagrani and L.~Samushia, \emph{{Information Content of the Angular
  Multipoles of Redshift-Space Galaxy Bispectrum}},
  \href{https://doi.org/10.1093/mnras/stx135}{\emph{Mon. Not. Roy. Astron.
  Soc.} {\bfseries 467} (2017) 928}
  [\href{https://arxiv.org/abs/1610.03488}{{\ttfamily 1610.03488}}].

\bibitem{Yankelevich2018}
V.~Yankelevich and C.~Porciani, \emph{{Cosmological information in the
  redshift-space bispectrum}},
  \href{https://doi.org/10.1093/mnras/sty3143}{\emph{Mon. Not. Roy. Astron.
  Soc.} {\bfseries 483} (2019) 2078}
  [\href{https://arxiv.org/abs/1807.07076}{{\ttfamily 1807.07076}}].

\bibitem{Mead2020}
A.~Mead, S.~Brieden, T.~Tr\"oster and C.~Heymans, \emph{{HMcode-2020: Improved
  modelling of non-linear cosmological power spectra with baryonic feedback}},
  \href{https://arxiv.org/abs/2009.01858}{{\ttfamily 2009.01858}}.

\bibitem{Smith2003}
R.~E. {Smith}, J.~A. {Peacock}, A.~{Jenkins}, S.~D.~M. {White}, C.~S. {Frenk},
  F.~R. {Pearce} et~al., \emph{{Stable clustering, the halo model and
  non-linear cosmological power spectra}},
  \href{https://doi.org/10.1046/j.1365-8711.2003.06503.x}{\emph{\mnras}
  {\bfseries 341} (2003) 1311}
  [\href{https://arxiv.org/abs/astro-ph/0207664}{{\ttfamily
  astro-ph/0207664}}].

\bibitem{Takahashi2012}
R.~{Takahashi}, M.~{Sato}, T.~{Nishimichi}, A.~{Taruya} and M.~{Oguri},
  \emph{{Revising the Halofit Model for the Nonlinear Matter Power Spectrum}},
  \href{https://doi.org/10.1088/0004-637X/761/2/152}{\emph{\apj} {\bfseries
  761} (2012) 152} [\href{https://arxiv.org/abs/1208.2701}{{\ttfamily
  1208.2701}}].

\bibitem{Smith:2018zcj}
R.~E. Smith and R.~E. Angulo, \emph{{Precision modelling of the matter power
  spectrum in a Planck-like Universe}},
  \href{https://doi.org/10.1093/mnras/stz890}{\emph{Mon. Not. Roy. Astron.
  Soc.} {\bfseries 486} (2019) 1448}
  [\href{https://arxiv.org/abs/1807.00040}{{\ttfamily 1807.00040}}].

\bibitem{Reimberg2018}
P.~Reimberg, F.~Bernardeau, T.~Nishimichi and M.~Rizzato, \emph{{Failures of
  Halofit model for computation of Fisher Matrices}},
  \href{https://arxiv.org/abs/1811.02976}{{\ttfamily 1811.02976}}.

\bibitem{Gil-Marin:2014sta}
H.~Gil-Mar{\'\i}n, J.~Nore{\~n}a, L.~Verde, W.~J. Percival, C.~Wagner,
  M.~Manera et~al., \emph{{The power spectrum and bispectrum of SDSS DR11 BOSS
  galaxies -- I. Bias and gravity}},
  \href{https://doi.org/10.1093/mnras/stv961}{\emph{Mon. Not. Roy. Astron.
  Soc.} {\bfseries 451} (2015) 539}
  [\href{https://arxiv.org/abs/1407.5668}{{\ttfamily 1407.5668}}].

\bibitem{Lazanu2015b}
A.~{Lazanu}, T.~{Giannantonio}, M.~{Schmittfull} and E.~P.~S. {Shellard},
  \emph{{Matter bispectrum of large-scale structure with Gaussian and
  non-Gaussian initial conditions: Halo models, perturbation theory, and a
  three-shape model}},
  \href{https://doi.org/10.1103/PhysRevD.95.083511}{\emph{\prd} {\bfseries 95}
  (2017) 083511} [\href{https://arxiv.org/abs/1511.02022}{{\ttfamily
  1511.02022}}].

\bibitem{Hashimoto:2017klo}
I.~Hashimoto, Y.~Rasera and A.~Taruya, \emph{{Precision cosmology with
  redshift-space bispectrum: a perturbation theory based model at one-loop
  order}}, \href{https://doi.org/10.1103/PhysRevD.96.043526}{\emph{Phys. Rev.
  D} {\bfseries 96} (2017) 043526}
  [\href{https://arxiv.org/abs/1705.02574}{{\ttfamily 1705.02574}}].

\bibitem{Chan2017}
K.~C. {Chan} and L.~{Blot}, \emph{{Assessment of the information content of the
  power spectrum and bispectrum}},
  \href{https://doi.org/10.1103/PhysRevD.96.023528}{\emph{\prd} {\bfseries 96}
  (2017) 023528} [\href{https://arxiv.org/abs/1610.06585}{{\ttfamily
  1610.06585}}].

\bibitem{Oddo:2019run}
A.~Oddo, E.~Sefusatti, C.~Porciani, P.~Monaco and A.~G. S{\'a}nchez,
  \emph{{Toward a robust inference method for the galaxy bispectrum: likelihood
  function and model selection}},
  \href{https://doi.org/10.1088/1475-7516/2020/03/056}{\emph{JCAP} {\bfseries
  03} (2020) 056} [\href{https://arxiv.org/abs/1908.01774}{{\ttfamily
  1908.01774}}].

\bibitem{Agarwal:2020lov}
N.~Agarwal, V.~Desjacques, D.~Jeong and F.~Schmidt, \emph{{Information content
  in the redshift-space galaxy power spectrum and bispectrum}},
  \href{https://arxiv.org/abs/2007.04340}{{\ttfamily 2007.04340}}.

\bibitem{MoradinezhadDizgah2020}
A.~Moradinezhad~Dizgah, M.~Biagetti, E.~Sefusatti, V.~Desjacques and
  J.~Nore\~na, \emph{{Primordial Non-Gaussianity from Biased Tracers:
  Likelihood Analysis of Real-Space Power Spectrum and Bispectrum}},
  \href{https://arxiv.org/abs/2010.14523}{{\ttfamily 2010.14523}}.

\bibitem{Alcock1979}
C.~{Alcock} and B.~{Paczynski}, \emph{{An evolution free test for non-zero
  cosmological constant}}, \href{https://doi.org/10.1038/281358a0}{\emph{\nat}
  {\bfseries 281} (1979) 358}.

\bibitem{Seo2003}
H.-J. Seo and D.~J. Eisenstein, \emph{{Probing dark energy with baryonic
  acoustic oscillations from future large galaxy redshift surveys}},
  \href{https://doi.org/10.1086/379122}{\emph{Astrophys. J.} {\bfseries 598}
  (2003) 720} [\href{https://arxiv.org/abs/astro-ph/0307460}{{\ttfamily
  astro-ph/0307460}}].

\bibitem{Song2015}
Y.-S. {Song}, A.~{Taruya} and A.~{Oka}, \emph{{Cosmology with anisotropic
  galaxy clustering from the combination of power spectrum and bispectrum}},
  \href{https://doi.org/10.1088/1475-7516/2015/08/007}{\emph{\jcap} {\bfseries
  8} (2015) 007} [\href{https://arxiv.org/abs/1502.03099}{{\ttfamily
  1502.03099}}].

\bibitem{2011ApJ...740L..20G}
Y.~{Gong}, X.~{Chen}, M.~{Silva}, A.~{Cooray} and M.~G. {Santos}, \emph{{The OH
  Line Contamination of 21 cm Intensity Fluctuation Measurements for z = 1-4}},
  \href{https://doi.org/10.1088/2041-8205/740/1/L20}{\emph{\apjl} {\bfseries
  740} (2011) L20} [\href{https://arxiv.org/abs/1108.0947}{{\ttfamily
  1108.0947}}].

\bibitem{Zaldarriaga2003b}
M.~{Zaldarriaga}, S.~R. {Furlanetto} and L.~{Hernquist}, \emph{{21 Centimeter
  Fluctuations from Cosmic Gas at High Redshifts}},
  \href{https://doi.org/10.1086/386327}{\emph{\apj} {\bfseries 608} (2004) 622}
  [\href{https://arxiv.org/abs/astro-ph/0311514}{{\ttfamily
  astro-ph/0311514}}].

\bibitem{Tegmark2008}
M.~{Tegmark} and M.~{Zaldarriaga}, \emph{{Fast Fourier transform telescope}},
  \href{https://doi.org/10.1103/PhysRevD.79.083530}{\emph{\prd} {\bfseries 79}
  (2009) 083530} [\href{https://arxiv.org/abs/0805.4414}{{\ttfamily
  0805.4414}}].

\bibitem{PUMA2018}
{Cosmic Visions 21 cm Collaboration}, R.~{Ansari}, E.~J. {Arena}, K.~{Bandura},
  P.~{Bull}, E.~{Castorina} et~al., \emph{{Inflation and Early Dark Energy with
  a Stage II Hydrogen Intensity Mapping Experiment}}, {\emph{arXiv e-prints}
  (2018) arXiv:1810.09572} [\href{https://arxiv.org/abs/1810.09572}{{\ttfamily
  1810.09572}}].

\bibitem{Santos:2017qgq}
{\scshape MeerKLASS} collaboration, M.~G. Santos et~al., \emph{{MeerKLASS:
  MeerKAT Large Area Synoptic Survey}},  in \emph{{MeerKAT Science}: {On the
  Pathway to the SKA}}, 9, 2017,
  \href{https://arxiv.org/abs/1709.06099}{{\ttfamily 1709.06099}}.

\bibitem{Fonseca:2019qek}
J.~Fonseca, J.-A. Viljoen and R.~Maartens, \emph{{Constraints on the growth
  rate using the observed galaxy power spectrum}},
  \href{https://doi.org/10.1088/1475-7516/2019/12/028}{\emph{JCAP} {\bfseries
  12} (2019) 028} [\href{https://arxiv.org/abs/1907.02975}{{\ttfamily
  1907.02975}}].

\bibitem{Shaw:2013wza}
J.~Shaw, K.~Sigurdson, U.-L. Pen, A.~Stebbins and M.~Sitwell, \emph{{All-Sky
  Interferometry with Spherical Harmonic Transit Telescopes}},
  \href{https://doi.org/10.1088/0004-637X/781/2/57}{\emph{Astrophys. J.}
  {\bfseries 781} (2014) 57} [\href{https://arxiv.org/abs/1302.0327}{{\ttfamily
  1302.0327}}].

\bibitem{Shaw:2014khi}
J.~Shaw, K.~Sigurdson, M.~Sitwell, A.~Stebbins and U.-L. Pen, \emph{{Coaxing
  cosmic 21 cm fluctuations from the polarized sky using m-mode analysis}},
  \href{https://doi.org/10.1103/PhysRevD.91.083514}{\emph{Phys. Rev. D}
  {\bfseries 91} (2015) 083514}
  [\href{https://arxiv.org/abs/1401.2095}{{\ttfamily 1401.2095}}].

\bibitem{Pober:2014lva}
J.~C. Pober, \emph{{The Impact of Foregrounds on Redshift Space Distortion
  Measurements With the Highly-Redshifted 21 cm Line}},
  \href{https://doi.org/10.1093/mnras/stu2575}{\emph{Mon. Not. Roy. Astron.
  Soc.} {\bfseries 447} (2015) 1705}
  [\href{https://arxiv.org/abs/1411.2050}{{\ttfamily 1411.2050}}].

\bibitem{Byrne:2018dkh}
R.~Byrne, M.~F. Morales, B.~Hazelton, W.~Li, N.~Barry, A.~P. Beardsley et~al.,
  \emph{{Fundamental Limitations on the Calibration of Redundant 21\,cm
  Cosmology Instruments and Implications for HERA and the SKA}},
  \href{https://doi.org/10.3847/1538-4357/ab107d}{\emph{Astrophys. J.}
  {\bfseries 875} (2019) 70}
  [\href{https://arxiv.org/abs/1811.01378}{{\ttfamily 1811.01378}}].

\bibitem{Jacobson:2003wv}
T.~Jacobson and R.~Parentani, \emph{{Horizon entropy}},
  \href{https://doi.org/10.1023/A:1023785123428}{\emph{Found. Phys.} {\bfseries
  33} (2003) 323} [\href{https://arxiv.org/abs/gr-qc/0302099}{{\ttfamily
  gr-qc/0302099}}].

\bibitem{Furlanetto:2006jb}
S.~Furlanetto, S.~Oh and F.~Briggs, \emph{{Cosmology at Low Frequencies: The
  21\,cm Transition and the High-Redshift Universe}},
  \href{https://doi.org/10.1016/j.physrep.2006.08.002}{\emph{Phys. Rept.}
  {\bfseries 433} (2006) 181}
  [\href{https://arxiv.org/abs/astro-ph/0608032}{{\ttfamily
  astro-ph/0608032}}].

\bibitem{Chang2007}
T.-C. Chang, U.-L. Pen, J.~B. Peterson and P.~McDonald, \emph{{Baryon Acoustic
  Oscillation Intensity Mapping as a Test of Dark Energy}},
  \href{https://doi.org/10.1103/PhysRevLett.100.091303}{\emph{Phys. Rev. Lett.}
  {\bfseries 100} (2008) 091303}
  [\href{https://arxiv.org/abs/0709.3672}{{\ttfamily 0709.3672}}].

\bibitem{Liu2011}
A.~{Liu} and M.~{Tegmark}, \emph{{A method for 21 cm power spectrum estimation
  in the presence of foregrounds}},
  \href{https://doi.org/10.1103/PhysRevD.83.103006}{\emph{\prd} {\bfseries 83}
  (2011) 103006} [\href{https://arxiv.org/abs/1103.0281}{{\ttfamily
  1103.0281}}].

\bibitem{Liu2012}
A.~{Liu} and M.~{Tegmark}, \emph{{How well can we measure and understand
  foregrounds with 21-cm experiments?}},
  \href{https://doi.org/10.1111/j.1365-2966.2011.19989.x}{\emph{\mnras}
  {\bfseries 419} (2012) 3491}
  [\href{https://arxiv.org/abs/1106.0007}{{\ttfamily 1106.0007}}].

\bibitem{Zhu:2016esh}
H.-M. Zhu, U.-L. Pen, Y.~Yu and X.~Chen, \emph{{Recovering lost 21 cm radial
  modes via cosmic tidal reconstruction}},
  \href{https://doi.org/10.1103/PhysRevD.98.043511}{\emph{Phys. Rev. D}
  {\bfseries 98} (2018) 043511}
  [\href{https://arxiv.org/abs/1610.07062}{{\ttfamily 1610.07062}}].

\bibitem{Karacayli:2019iyd}
N.~G. Kara{\c c}ayl\i and N.~Padmanabhan, \emph{{Anatomy of Cosmic Tidal
  Reconstruction}}, \href{https://doi.org/10.1093/mnras/stz964}{\emph{Mon. Not.
  Roy. Astron. Soc.} {\bfseries 486} (2019) 3864}
  [\href{https://arxiv.org/abs/1904.01387}{{\ttfamily 1904.01387}}].

\bibitem{Modi:2019hnu}
C.~Modi, M.~White, A.~Slosar and E.~Castorina, \emph{{Reconstructing
  large-scale structure with neutral hydrogen surveys}},
  \href{https://doi.org/10.1088/1475-7516/2019/11/023}{\emph{JCAP} {\bfseries
  11} (2019) 023} [\href{https://arxiv.org/abs/1907.02330}{{\ttfamily
  1907.02330}}].

\bibitem{Jasche2010}
J.~{Jasche} and F.~S. {Kitaura}, \emph{{Fast Hamiltonian sampling for
  large-scale structure inference}},
  \href{https://doi.org/10.1111/j.1365-2966.2010.16897.x}{\emph{\mnras}
  {\bfseries 407} (2010) 29} [\href{https://arxiv.org/abs/0911.2496}{{\ttfamily
  0911.2496}}].

\bibitem{Kitaura2013}
F.~S. {Kitaura}, \emph{{The initial conditions of the universe from constrained
  simulations.}}, \href{https://doi.org/10.1093/mnrasl/sls029}{\emph{\mnras}
  {\bfseries 429} (2013) L84}
  [\href{https://arxiv.org/abs/1203.4184}{{\ttfamily 1203.4184}}].

\bibitem{Wang:2014hia}
H.~Wang, H.~Mo, X.~Yang, Y.~Jing and W.~Lin, \emph{{ELUCID - Exploring the
  Local Universe with Reconstructed Initial Density field I: Hamiltonian Markov
  Chain Monte Carlo Method with Particle Mesh Dynamics}},
  \href{https://doi.org/10.1088/0004-637X/794/1/94}{\emph{Astrophys. J.}
  {\bfseries 794} (2014) 94} [\href{https://arxiv.org/abs/1407.3451}{{\ttfamily
  1407.3451}}].

\bibitem{Jasche:2014vpa}
J.~Jasche, F.~Leclercq and B.~D. Wandelt, \emph{{Past and present cosmic
  structure in the SDSS DR7 main sample}},
  \href{https://doi.org/10.1088/1475-7516/2015/01/036}{\emph{JCAP} {\bfseries
  01} (2015) 036} [\href{https://arxiv.org/abs/1409.6308}{{\ttfamily
  1409.6308}}].

\bibitem{Wang:2016qbz}
H.~Wang, H.~Mo, X.~Yang, Y.~Zhang, J.~Shi, Y.~Jing et~al., \emph{{ELUCID -
  Exploring the Local Universe with reConstructed Initial Density field III:
  Constrained Simulation in the SDSS Volume}},
  \href{https://doi.org/10.3847/0004-637X/831/2/164}{\emph{Astrophys. J.}
  {\bfseries 831} (2016) 164}
  [\href{https://arxiv.org/abs/1608.01763}{{\ttfamily 1608.01763}}].

\bibitem{Seljak:2017rmr}
U.~Seljak, G.~Aslanyan, Y.~Feng and C.~Modi, \emph{{Towards optimal extraction
  of cosmological information from nonlinear data}},
  \href{https://doi.org/10.1088/1475-7516/2017/12/009}{\emph{JCAP} {\bfseries
  12} (2017) 009} [\href{https://arxiv.org/abs/1706.06645}{{\ttfamily
  1706.06645}}].

\bibitem{Modi:2018cfi}
C.~Modi, Y.~Feng and U.~Seljak, \emph{{Cosmological Reconstruction From Galaxy
  Light: Neural Network Based Light-Matter Connection}},
  \href{https://doi.org/10.1088/1475-7516/2018/10/028}{\emph{JCAP} {\bfseries
  10} (2018) 028} [\href{https://arxiv.org/abs/1805.02247}{{\ttfamily
  1805.02247}}].

\bibitem{Parsons2012}
A.~R. {Parsons}, J.~C. {Pober}, J.~E. {Aguirre}, C.~L. {Carilli}, D.~C.
  {Jacobs} and D.~F. {Moore}, \emph{{A Per-baseline, Delay-spectrum Technique
  for Accessing the 21 cm Cosmic Reionization Signature}},
  \href{https://doi.org/10.1088/0004-637X/756/2/165}{\emph{\apj} {\bfseries
  756} (2012) 165} [\href{https://arxiv.org/abs/1204.4749}{{\ttfamily
  1204.4749}}].

\bibitem{Pober2014}
J.~C. {Pober}, A.~{Liu}, J.~S. {Dillon}, J.~E. {Aguirre}, J.~D. {Bowman}, R.~F.
  {Bradley} et~al., \emph{{What Next-generation 21 cm Power Spectrum
  Measurements can Teach us About the Epoch of Reionization}},
  \href{https://doi.org/10.1088/0004-637X/782/2/66}{\emph{\apj} {\bfseries 782}
  (2014) 66} [\href{https://arxiv.org/abs/1310.7031}{{\ttfamily 1310.7031}}].

\bibitem{Seo2015}
H.-J. Seo and C.~M. Hirata, \emph{{The foreground wedge and 21 cm BAO
  surveys}}, \href{https://doi.org/10.1093/mnras/stv2806}{\emph{Mon. Not. Roy.
  Astron. Soc.} {\bfseries 456} (2016) 3142}
  [\href{https://arxiv.org/abs/1508.06503}{{\ttfamily 1508.06503}}].

\bibitem{Pober2015}
J.~C. {Pober}, \emph{{The impact of foregrounds on redshift space distortion
  measurements with the highly redshifted 21-cm line}},
  \href{https://doi.org/10.1093/mnras/stu2575}{\emph{\mnras} {\bfseries 447}
  (2015) 1705} [\href{https://arxiv.org/abs/1411.2050}{{\ttfamily 1411.2050}}].

\bibitem{Fornazier:2021ini}
K.~S.~F. Fornazier et~al., \emph{{The BINGO Project V: Further steps in
  Component Separation and Bispectrum Analysis}},
  \href{https://doi.org/10.1051/0004-6361/202141707}{\emph{Astron. Astrophys.}
  {\bfseries 664} (2022) A18}
  [\href{https://arxiv.org/abs/2107.01637}{{\ttfamily 2107.01637}}].

\bibitem{Wang:2020lkn}
J.~Wang et~al., \emph{{HI intensity mapping with MeerKAT: calibration pipeline
  for multidish autocorrelation observations}},
  \href{https://doi.org/10.1093/mnras/stab1365}{\emph{Mon. Not. Roy. Astron.
  Soc.} {\bfseries 505} (2021) 3698}
  [\href{https://arxiv.org/abs/2011.13789}{{\ttfamily 2011.13789}}].

\bibitem{Li:2020bcr}
Y.~Li, M.~G. Santos, K.~Grainge, S.~Harper and J.~Wang, \emph{{HI intensity
  mapping with MeerKAT: 1/f noise analysis}},
  \href{https://doi.org/10.1093/mnras/staa3856}{\emph{Mon. Not. Roy. Astron.
  Soc.} {\bfseries 501} (2021) 4344}
  [\href{https://arxiv.org/abs/2007.01767}{{\ttfamily 2007.01767}}].

\bibitem{Matshawule:2020fjz}
S.~D. Matshawule, M.~Spinelli, M.~G. Santos and S.~Ngobese, \emph{{HI intensity
  mapping with MeerKAT: primary beam effects on foreground cleaning}},
  \href{https://doi.org/10.1093/mnras/stab1688}{\emph{Mon. Not. Roy. Astron.
  Soc.} {\bfseries 506} (2021) 5075}
  [\href{https://arxiv.org/abs/2011.10815}{{\ttfamily 2011.10815}}].

\bibitem{Liu:2019awk}
A.~Liu and J.~R. Shaw, \emph{{Data Analysis for Precision 21 cm Cosmology}},
  \href{https://doi.org/10.1088/1538-3873/ab5bfd}{\emph{Publ. Astron. Soc.
  Pac.} {\bfseries 132} (2020) 062001}
  [\href{https://arxiv.org/abs/1907.08211}{{\ttfamily 1907.08211}}].

\bibitem{Tegmark1997}
M.~Tegmark, \emph{{Measuring cosmological parameters with galaxy surveys}},
  \href{https://doi.org/10.1103/PhysRevLett.79.3806}{\emph{Phys. Rev. Lett.}
  {\bfseries 79} (1997) 3806}
  [\href{https://arxiv.org/abs/astro-ph/9706198}{{\ttfamily
  astro-ph/9706198}}].

\bibitem{Seo_2003}
H.-J. {Seo} and D.~J. {Eisenstein}, \emph{{Probing Dark Energy with Baryonic
  Acoustic Oscillations from Future Large Galaxy Redshift Surveys}},
  \href{https://doi.org/10.1086/379122}{\emph{\apj} {\bfseries 598} (2003) 720}
  [\href{https://arxiv.org/abs/astro-ph/0307460}{{\ttfamily
  astro-ph/0307460}}].

\bibitem{Lahav1991}
O.~Lahav, P.~B. Lilje, J.~R. Primack and M.~J. Rees, \emph{{Dynamical effects
  of the cosmological constant}}, {\emph{Mon. Not. Roy. Astron. Soc.}
  {\bfseries 251} (1991) 128}.

\bibitem{Linder2005}
E.~V. Linder, \emph{{Cosmic growth history and expansion history}},
  \href{https://doi.org/10.1103/PhysRevD.72.043529}{\emph{Phys. Rev. D}
  {\bfseries 72} (2005) 043529}
  [\href{https://arxiv.org/abs/astro-ph/0507263}{{\ttfamily
  astro-ph/0507263}}].

\bibitem{Albrecht:2006um}
A.~Albrecht et~al., \emph{{Report of the Dark Energy Task Force}},
  \href{https://arxiv.org/abs/astro-ph/0609591}{{\ttfamily astro-ph/0609591}}.

\bibitem{Sefusatti2006}
E.~{Sefusatti}, M.~{Crocce}, S.~{Pueblas} and R.~{Scoccimarro},
  \emph{{Cosmology and the bispectrum}},
  \href{https://doi.org/10.1103/PhysRevD.74.023522}{\emph{\prd} {\bfseries 74}
  (2006) 023522} [\href{https://arxiv.org/abs/astro-ph/0604505}{{\ttfamily
  astro-ph/0604505}}].

\bibitem{Sefusatti2007}
E.~{Sefusatti} and E.~{Komatsu}, \emph{{Bispectrum of galaxies from
  high-redshift galaxy surveys: Primordial non-Gaussianity and nonlinear galaxy
  bias}}, \href{https://doi.org/10.1103/PhysRevD.76.083004}{\emph{\prd}
  {\bfseries 76} (2007) 083004}
  [\href{https://arxiv.org/abs/0705.0343}{{\ttfamily 0705.0343}}].

\bibitem{Howlett:2017vwp}
C.~Howlett and W.~J. Percival, \emph{{Galaxy two-point covariance matrix
  estimation for next generation surveys}},
  \href{https://doi.org/10.1093/mnras/stx2342}{\emph{Mon. Not. Roy. Astron.
  Soc.} {\bfseries 472} (2017) 4935}
  [\href{https://arxiv.org/abs/1709.03057}{{\ttfamily 1709.03057}}].

\bibitem{Barreira:2017kxd}
A.~Barreira and F.~Schmidt, \emph{{Response Approach to the Matter Power
  Spectrum Covariance}},
  \href{https://doi.org/10.1088/1475-7516/2017/11/051}{\emph{JCAP} {\bfseries
  11} (2017) 051} [\href{https://arxiv.org/abs/1705.01092}{{\ttfamily
  1705.01092}}].

\bibitem{Li:2018scc}
Y.~Li, S.~Singh, B.~Yu, Y.~Feng and U.~Seljak, \emph{{Disconnected Covariance
  of 2-point Functions in Large-Scale Structure}},
  \href{https://doi.org/10.1088/1475-7516/2019/01/016}{\emph{JCAP} {\bfseries
  01} (2019) 016} [\href{https://arxiv.org/abs/1811.05714}{{\ttfamily
  1811.05714}}].

\bibitem{Blot:2018oxk}
L.~Blot et~al., \emph{{Comparing approximate methods for mock catalogues and
  covariance matrices II: Power spectrum multipoles}},
  \href{https://doi.org/10.1093/mnras/stz507}{\emph{Mon. Not. Roy. Astron.
  Soc.} {\bfseries 485} (2019) 2806}
  [\href{https://arxiv.org/abs/1806.09497}{{\ttfamily 1806.09497}}].

\bibitem{Gualdi:2020ymf}
D.~Gualdi and L.~Verde, \emph{{Galaxy redshift-space bispectrum: the Importance
  of Being Anisotropic}},
  \href{https://doi.org/10.1088/1475-7516/2020/06/041}{\emph{JCAP} {\bfseries
  06} (2020) 041} [\href{https://arxiv.org/abs/2003.12075}{{\ttfamily
  2003.12075}}].

\bibitem{Biagetti:2021tua}
M.~Biagetti, L.~Castiblanco, J.~Nore\~na and E.~Sefusatti, \emph{{The
  covariance of squeezed bispectrum configurations}},
  \href{https://doi.org/10.1088/1475-7516/2022/09/009}{\emph{JCAP} {\bfseries
  09} (2022) 009} [\href{https://arxiv.org/abs/2111.05887}{{\ttfamily
  2111.05887}}].

\bibitem{Floss:2022wkq}
T.~Fl\"oss, M.~Biagetti and P.~D. Meerburg, \emph{{Primordial non-Gaussianity
  and non-Gaussian Covariance}},
  \href{https://arxiv.org/abs/2206.10458}{{\ttfamily 2206.10458}}.

\bibitem{Hahn:2019zob}
C.~Hahn, F.~Villaescusa-Navarro, E.~Castorina and R.~Scoccimarro,
  \emph{{Constraining $M_\nu$ with the bispectrum. Part I. Breaking parameter
  degeneracies}},
  \href{https://doi.org/10.1088/1475-7516/2020/03/040}{\emph{JCAP} {\bfseries
  03} (2020) 040} [\href{https://arxiv.org/abs/1909.11107}{{\ttfamily
  1909.11107}}].

\bibitem{Hahn:2020lou}
C.~Hahn and F.~Villaescusa-Navarro, \emph{{Constraining $M_\nu$ with the
  bispectrum. Part II. The information content of the galaxy bispectrum
  monopole}}, \href{https://doi.org/10.1088/1475-7516/2021/04/029}{\emph{JCAP}
  {\bfseries 04} (2021) 029}
  [\href{https://arxiv.org/abs/2012.02200}{{\ttfamily 2012.02200}}].

\bibitem{Baldauf2016}
T.~{Baldauf}, M.~{Mirbabayi}, M.~{Simonovi{\'c}} and M.~{Zaldarriaga},
  \emph{{LSS constraints with controlled theoretical uncertainties}},
  {\emph{ArXiv e-prints} (2016) }
  [\href{https://arxiv.org/abs/1602.00674}{{\ttfamily 1602.00674}}].

\bibitem{Baldauf:2015aha}
T.~Baldauf, L.~Mercolli and M.~Zaldarriaga, \emph{{Effective field theory of
  large scale structure at two loops: The apparent scale dependence of the
  speed of sound}},
  \href{https://doi.org/10.1103/PhysRevD.92.123007}{\emph{Phys. Rev. D}
  {\bfseries 92} (2015) 123007}
  [\href{https://arxiv.org/abs/1507.02256}{{\ttfamily 1507.02256}}].

\bibitem{Chudaykin:2019ock}
A.~Chudaykin and M.~M. Ivanov, \emph{{Measuring neutrino masses with
  large-scale structure: Euclid forecast with controlled theoretical error}},
  \href{https://doi.org/10.1088/1475-7516/2019/11/034}{\emph{JCAP} {\bfseries
  11} (2019) 034} [\href{https://arxiv.org/abs/1907.06666}{{\ttfamily
  1907.06666}}].

\bibitem{GilMarin2014}
H.~{Gil-Mar{\'{\i}}n}, C.~{Wagner}, J.~{Nore{\~n}a}, L.~{Verde} and
  W.~{Percival}, \emph{{Dark matter and halo bispectrum in redshift space:
  theory and applications}},
  \href{https://doi.org/10.1088/1475-7516/2014/12/029}{\emph{\jcap} {\bfseries
  12} (2014) 029} [\href{https://arxiv.org/abs/1407.1836}{{\ttfamily
  1407.1836}}].

\bibitem{GilMarin2017}
H.~{Gil-Mar{\'{\i}}n}, W.~J. {Percival}, L.~{Verde}, J.~R. {Brownstein}, C.-H.
  {Chuang}, F.-S. {Kitaura} et~al., \emph{{The clustering of galaxies in the
  SDSS-III Baryon Oscillation Spectroscopic Survey: RSD measurement from the
  power spectrum and bispectrum of the DR12 BOSS galaxies}},
  \href{https://doi.org/10.1093/mnras/stw2679}{\emph{\mnras} {\bfseries 465}
  (2017) 1757} [\href{https://arxiv.org/abs/1606.00439}{{\ttfamily
  1606.00439}}].

\bibitem{Karagiannis2018}
D.~{Karagiannis}, A.~{Lazanu}, M.~{Liguori}, A.~{Raccanelli}, N.~{Bartolo} and
  L.~{Verde}, \emph{{Constraining primordial non-Gaussianity with bispectrum
  and power spectrum from upcoming optical and radio surveys}},
  \href{https://doi.org/10.1093/mnras/sty1029}{\emph{\mnras} {\bfseries 478}
  (2018) 1341} [\href{https://arxiv.org/abs/1801.09280}{{\ttfamily
  1801.09280}}].

\bibitem{Planck:2015bue}
{\scshape Planck} collaboration, P.~A.~R. Ade et~al., \emph{{Planck 2015
  results. XIV. Dark energy and modified gravity}},
  \href{https://doi.org/10.1051/0004-6361/201525814}{\emph{Astron. Astrophys.}
  {\bfseries 594} (2016) A14}
  [\href{https://arxiv.org/abs/1502.01590}{{\ttfamily 1502.01590}}].

\bibitem{Joyce:2016vqv}
A.~Joyce, L.~Lombriser and F.~Schmidt, \emph{{Dark Energy Versus Modified
  Gravity}},
  \href{https://doi.org/10.1146/annurev-nucl-102115-044553}{\emph{Ann. Rev.
  Nucl. Part. Sci.} {\bfseries 66} (2016) 95}
  [\href{https://arxiv.org/abs/1601.06133}{{\ttfamily 1601.06133}}].

\bibitem{Amendola:2016saw}
L.~Amendola et~al., \emph{{Cosmology and fundamental physics with the Euclid
  satellite}}, \href{https://doi.org/10.1007/s41114-017-0010-3}{\emph{Living
  Rev. Rel.} {\bfseries 21} (2018) 2}
  [\href{https://arxiv.org/abs/1606.00180}{{\ttfamily 1606.00180}}].

\bibitem{Slosar:2019flp}
A.~Slosar et~al., \emph{{Dark Energy and Modified Gravity}},
  \href{https://arxiv.org/abs/1903.12016}{{\ttfamily 1903.12016}}.

\bibitem{Frusciante:2019xia}
N.~Frusciante and L.~Perenon, \emph{{Effective field theory of dark energy: A
  review}}, \href{https://doi.org/10.1016/j.physrep.2020.02.004}{\emph{Phys.
  Rept.} {\bfseries 857} (2020) 1}
  [\href{https://arxiv.org/abs/1907.03150}{{\ttfamily 1907.03150}}].

\bibitem{DiValentino:2021izs}
E.~Di~Valentino, O.~Mena, S.~Pan, L.~Visinelli, W.~Yang, A.~Melchiorri et~al.,
  \emph{{In the realm of the Hubble tension -- a review of solutions}},
  \href{https://doi.org/10.1088/1361-6382/ac086d}{\emph{Class. Quant. Grav.}
  {\bfseries 38} (2021) 153001}
  [\href{https://arxiv.org/abs/2103.01183}{{\ttfamily 2103.01183}}].

\bibitem{Cunnington:2022ryj}
S.~Cunnington, \emph{{Detecting the power spectrum turnover with HI intensity
  mapping}}, \href{https://doi.org/10.1093/mnras/stac576}{\emph{Mon. Not. Roy.
  Astron. Soc.} {\bfseries 512} (2022) 2408}
  [\href{https://arxiv.org/abs/2202.13828}{{\ttfamily 2202.13828}}].

\bibitem{Viljoen:2021ypp}
J.-A. Viljoen, J.~Fonseca and R.~Maartens, \emph{{Multi-wavelength
  spectroscopic probes: prospects for primordial non-Gaussianity and
  relativistic effects}},
  \href{https://doi.org/10.1088/1475-7516/2021/11/010}{\emph{JCAP} {\bfseries
  11} (2021) 010} [\href{https://arxiv.org/abs/2107.14057}{{\ttfamily
  2107.14057}}].

\bibitem{Paul:2020ank}
S.~Paul, M.~G. Santos, J.~Townsend, M.~J. Jarvis, N.~Maddox, J.~D. Collier
  et~al., \emph{{HI intensity mapping with the MIGHTEE survey: power spectrum
  estimates}}, \href{https://doi.org/10.1093/mnras/stab1089}{\emph{Mon. Not.
  Roy. Astron. Soc.} {\bfseries 505} (2021) 2039}
  [\href{https://arxiv.org/abs/2009.13550}{{\ttfamily 2009.13550}}].

\bibitem{Cunnington:2022uzo}
S.~Cunnington et~al., \emph{{HI intensity mapping with MeerKAT: power spectrum
  detection in cross-correlation with WiggleZ galaxies}},
  \href{https://arxiv.org/abs/2206.01579}{{\ttfamily 2206.01579}}.

\bibitem{Jolicoeur:2020eup}
S.~Jolicoeur, R.~Maartens, E.~M. De~Weerd, O.~Umeh, C.~Clarkson and S.~Camera,
  \emph{{Detecting the relativistic bispectrum in 21cm intensity maps}},
  \href{https://doi.org/10.1088/1475-7516/2021/06/039}{\emph{JCAP} {\bfseries
  06} (2021) 039} [\href{https://arxiv.org/abs/2009.06197}{{\ttfamily
  2009.06197}}].

\end{thebibliography}\endgroup

\end{document}